\documentclass[aps,preprint,floatfix,nofootinbib,showpacs]{revtex4}
\usepackage{graphicx}

\newcommand{\U}{{\cal U}}
\begin{document}

%\begin{flushright}}
%\preprint{hep-ph/yymmnnn}
%\end{flushright}

%\bibliographystyle{revtex}

\title{Collider Phenomenology of Unparticle Physics}

\renewcommand{\thefootnote}{\fnsymbol{footnote}}

\author{ 
Kingman Cheung$^{1,2}$, 
Wai-Yee Keung$^3$
and Tzu-Chiang Yuan$^1$
 }
\affiliation{$^1$Department of Physics, National Tsing Hua University, 
Hsinchu 300
\\
$^2$Physics Division, National Center for Theoretical Sciences,
Hsinchu 300
\\
$^3$Department of Physics, University of Illinois, Chicago IL 60628
}

\renewcommand{\thefootnote}{\arabic{footnote}}
\date{\today}

\begin{abstract}
Low energy phenomenology of the unparticle physics associated with an
exact scale invariant sector possessing a non-trivial infrared fixed point at a
higher energy scale is explored for both electron-positron and 
hadronic colliders.  
Feynman rules for a spin 0, 1 or 2
unparticle coupled to a variety of standard model gauge invariant
operators that are relevant to many low energy processes involving
either real emissions of unparticles or their virtual propagator
effects are presented.  Missing energy and/or recoil mass
distributions of the unparticle in the associated production of
unparticle together with a photon or $Z$ boson at LEP2 and ILC as well
as in $Z$ decay into an unparticle plus a fermion-antifermion pair are
studied.  In addition, mono-jet production with
missing energy from the unparticle at hadronic collisions are
explored.  The complex phase in the unparticle propagator that can
give rise to interesting interference effects between an unparticle
exchange diagram and the standard model amplitudes are studied in
details for the Drell-Yan process as well as muon pair and diphoton
production in electron-positron annihilation.  These energy and/or
recoil mass distributions (with the exception in hadron colliders) and
interference effects are found sensitively depending not only on the
scale dimension but also on the spin of the unparticle.  For the spin-2
unparticle, its physical effects is found to resemble that of a tower
of Kaluza-Klein gravitons, which strongly indicates that the underlying 
unparticle physics may have root in a higher dimensional theory.  
A connection between unparticle physics and theories of large extra dimension 
is speculated.
Experimental constraints on the unparticle scale are deduced from the 
LEP2 data on mono-photon production and from 
the 4-fermion contact interactions.
\end{abstract}


\maketitle
 
\section{Introduction}

Scale invariance is a very appealing symmetry in both physics and
mathematics.  The dilatation generator $D$ for scale transformation
does not commute with the spacetime translation generators
$P_\mu$. Their commutation relations are familiar:
\begin{equation}
[D,P_\mu] = - i P_\mu \; .
\end{equation}
This implies for a real $s$
\begin{equation}
\exp (+i s D) P^2 \exp(-i s D) = \exp(2 s) P^2 \; .
\end{equation}
Thus, the exact scale symmetry requires that the mass spectrum is either 
continuous or all masses are zero.
In a renormalizable theory, this symmetry must be broken either 
explicitly by some dimensional mass
parameters in the theory or implicitly by quantum loop effects, {\it \`{a} la} 
Coleman-Weinberg mechanism \cite{coleman-weinberg}, for example.
Indeed, scale invariance is manifestly broken in the Lagrangian of the
standard model (SM) of particle physics at tree level by just a single
negative mass squared term in the Higgs potential. Despite the lack of
scale invariance in the standard model, it is logically plausible to
imagine that there exists such a scale invariant sector at a higher
scale above TeV that can be probed at the LHC or ILC. Such a sector
might be strongly coupled to itself and highly nontrivial but
nevertheless can be only weakly coupled to the matter in the standard
model.  One expects that such a sector decouples effectively from the low
energy and can use the power of effective field theory approach to describe its
low energy effects.
 
Recently, Georgi \cite{georgi1} motivated by the Banks-Zaks theory
\cite{banks-zaks}, suggested that a scale invariant sector with a
nontrivial infrared fixed-point behaves rather peculiar from the
perspective of particle physics.  It was keenly observed in
\cite{georgi1} that an operator $\cal O_\U$ with a general
non-integral scale dimension $d_\U$ in a scale invariant sector has a
mass spectrum looked like a $d_{\cal U}$ number of invisible massless
particles.  This was coined as unparticle $\U$ by Georgi.  
Unparticle does not have a fixed invariant mass squared but instead a 
continuous mass spectrum in accordance with the above general argument.
It was also pointed out that real production of an unparticle at 
low energy processes
described by an effective field theory can give rise to peculiar
missing energy distributions because of the possible non-integral values of
$d_\U$.
%%%

Subsequently, the propagator for the unparticle was worked out independently 
in \cite{georgi2} and  \cite{ours}.  An unusual
phase in the unparticle propagator was discovered by both groups and
the interesting interference patterns between the amplitude of 
$s$-channel unparticle 
exchange and those from the SM were studied.  
In Ref.~\cite{georgi2}, the interference effect between the complex phase of
the unparticle propagator and the complex Breit-Wigner form of the 
unstable $Z$ boson propagator was studied in details for the backward-forward
asymmetry in the $e^-e^+ \to \mu^- \mu^+$ process near the $Z$ pole.
In Ref.~\cite{ours}, 
the interference between the amplitude of an $s$-channel spin-1 unparticle
exchange with the SM amplitudes for the Drell-Yan process
was explored at the Tevatron. An one-loop unparticle exchange contribution 
to the lepton anomalous magnetic moment was also calculated in \cite{ours}.
More recently, various phenomenology of the unparticle has been 
explored by many groups 
\cite{Luo-Zhu},
\cite{Chen-Geng},
\cite{Ding-Yan},
\cite{Liao},
\cite{Aliev-Cornell-Gaur},
\cite{Li-Wei},
\cite{Lu-Wang-Wang},
\cite{Fox-Rajaraman-Shirman},
\cite{Greiner},
\cite{Davoudiasl},
\cite{Choudhury-Ghosh-Mamta},
\cite{Chen-He},
\cite{Mathews-Ravindran},
\cite{Zhou},
\cite{Liao-Liu},
\cite{minimal-walking},
\cite{Bander-Feng-Rajaraman-Shirman},
\cite{Rizzo},
\cite{ungravity}.

In this paper, we present in much more details the results reported
earlier in \cite{ours} and extend to further processes that are relevant
to collider experiments.  We believe these processes are of immediate
interests to theoretical and experimental communities.
In the next section, we review the derivation of the two-point
functions \cite{georgi1}, propagators \cite{georgi2},\cite{ours} and spin
structures of the unparticle operators $O_\U$, $O^\mu_\U$ and
$O^{\mu\nu}_\U$ first introduced in Ref.\cite{georgi1}. Feynman rules
for these operators coupled to those standard model invariant operators
of special interests are explicitly given.  
In addition, four-fermion
contact 
interactions due to spin-1 and 2 unparticle exchanges are written
down.  At the end of this section, 
we also speculate on a possible connection between unparticle physics and 
theories of large extra dimension.
The subsequent two sections are phenomenological applications.
In section III, we discuss real emissions of
unparticles.  This covers $e^- e^+ \to \gamma \U$ and $e^- e^+ \to Z
\U$ at $e^-e^+$ colliders, and $Z \to f \bar f \U$ at the $Z$ pole, as
well as mono-jet 
production plus unparticle $\U$ at
hadron colliders.  LEP2 data of mono-photon production is used 
to constrain the 
unparticle scale.
In Sec. IV, we study the interference effects
between the exchange of virtual unparticle and the standard model
amplitudes.  We discuss several classic reactions including Drell-Yan
process, $e^- e^+ \to f \bar f$ with $f \neq e$ and $f \bar f \to
\gamma \gamma$.  Experimental limits of the 4-fermion contact interactions 
from global fits are also used to constrain the unparticle scale.
Conclusions and comments will be given in section V.
Some tedious formulas are relegated in an appendix.

\section{Formalism}

To fix notation we denote the scale invariant sector as a Banks-Zaks
($\cal BZ$) sector \cite{banks-zaks} and follow closely the scenario
studied in \cite{georgi1}.  The $\cal BZ$ sector can interact with the
standard model fields through the exchange of a connector sector that
has a high mass scale $M_\U$. Below this high mass scale,
non-renormalizable operators that are suppressed by inverse powers of
$M_\U$ are induced. Generically, we have operators of the form
\begin{equation}
   \frac{ 1 } {M_\U^{d_{\mathrm SM} + d_{\cal BZ}-4}} \, 
   {\cal O}_{\mathrm SM} \, {\cal O}_{\cal BZ}  \; ,
\label{genericop}
\end{equation}
where ${\cal O}_{\mathrm SM}$ and ${\cal O_{BZ}}$ represent local operators
constructed out of standard model and ${\cal BZ}$ fields 
with scale dimensions $d_{\mathrm SM}$ and $d_{\cal BZ}$,
respectively. As in massless non-abelian gauge theories,
renormalization effects in the scale invariant $\cal BZ$ sector induce
dimensional transmutation \cite{coleman-weinberg} at an energy scale
$\Lambda_\U$ . Below $\Lambda_\U$ matching conditions must be imposed
onto the operator (\ref{genericop}) to match a new set of operators
having the following form
\begin{equation}
 \label{effectiveop}
    C_{\cal O_U} \frac{ \Lambda_\U^{d_{\cal BZ} - d_\U} } {M^{d_{\mathrm SM} + d_{\cal BZ}-4}_\U } \,
   {\cal O}_{\mathrm SM}\, {\cal O}_\U \;,
\end{equation}
where  $d_\U$ is the scale dimension of the unparticle operator ${\cal O_U}$ and
$C_{\cal O_\U}$ is a coefficient function fixed by the matching.
Whether this matching can be implemented is highly
nontrivial since the scale invariant sector might be strongly coupled.
While we are very much ignorant of this scale invariant sector above
the TeV scale, it was argued in \cite{georgi1} that using the
effective field theory approach specified by the operators like
Eq.~(\ref{effectiveop}) one should be able to probe the unparticle
physics at the LHC and ILC.
Throughout this work, it is tacitly assumed that an exact scale invariance sector 
survives all the way down to the electroweak scale.

Three unparticle operators with different Lorentz structures were
addressed in \cite{georgi1}: $\left\{ O_\U, O^\mu_\U, O_\U^{\mu\nu}
\right\} \in {\cal O_U}$, which correspond to scalar, vector, and
tensor operators, respectively. 
Spin-$\frac{1}{2}$ unparticle operator was considered in \cite{Luo-Zhu}. 
In general, an unparticle operator from a scale invariant sector can
 be labeled by a triple 
$(d_\U ; j_1, j_2)$ where $d_\U$ is its scale dimension and $2 j_1$ 
and $2j_2$ are two
integers labeling the representation of the Lorentz group that it belongs to. 
Unitarity imposes constraints on possible values taken by the scale dimension
 depending on 
$j_1$ and $j_2$ \cite{Mack}.
For example, for the scalar unparticle operator $O_\U$, $j_1 = j_2 = 0$ and
 unitarity constrains $d_\U > 1$.
In the numerical works presented in this paper, we will simply require
 $d_\U > 1$ 
for all unparticle operators.
These unparticle operators  can even carry standard model
quantum numbers \cite{georgi1}, for example a charged
unparticle can be anticipated. 
Throughout this work we are contented with the unparticle operators that are 
standard model singlets.

\subsection{Phase space for real emission of unparticle}
It was demonstrated in \cite{georgi1} that  scale invariance can be 
used to fix the two-point functions of 
the unparticle operators. 
Let us consider a two-point function for a scalar unparticle operator $O_\U$
 \begin{eqnarray}
 \label{2ptfun}
  \langle 0 \vert O_\U (x)  O_\U^\dagger (0) \vert 0 \rangle 
&=& \langle 0| e^{i \hat P \cdot x}O_\U(0)e^{-i \hat P \cdot x}O_\U^\dagger(0)|0\rangle 
  \nonumber\\
   &=&  \int d \lambda \int d \lambda^\prime \langle 0| O_\U(0) \vert \lambda^\prime \rangle \langle \lambda^\prime \vert e^{-i \hat P \cdot x}
 \vert \lambda \rangle \langle \lambda \vert O_\U^\dagger(0)|0\rangle  \nonumber \\
&=& \int \frac{ d^4 P}{(2\pi)^4}   \, e^{-i P \cdot x} \, 
%\vert \langle 0 \vert O_\U (0)\vert P \rangle \vert^2
\rho_\U(P^2) \, ,
\end{eqnarray}
where $\rho_\U(P^2)$ is the spectral density and is formally given by
\begin{eqnarray}
\rho_\U(P^2)
& = & (2 \pi)^4 \int d \lambda \, \delta^4 (P - p_\lambda) \vert \langle 0 \vert O_\U(0) \vert \lambda \rangle \vert^2 \; .
\end{eqnarray}
%
%Note that the second line in Eq.~(\ref{2ptfun}) is obtained by 
%insertion of a complete set of states between operators. 
%
Inverse Fourier transformation gives
\begin{eqnarray}
%|\langle 0| O_\U(0)|P^2\rangle|^2
\rho_\U(P^2)
&=& \int d^4 x  \,  e^{i P \cdot x}\langle 0| O_\U(x)
   O_\U^\dagger(0)|0\rangle \nonumber \\
   & = & A_{d_\U} \ \theta(P^0) \ \theta(P^2) \ (P^2)^\alpha
\end{eqnarray}
where $\alpha$ is an index to be determined based on scale invariance
and $A_{d_\U}$ is a normalization factor also required to be fixed.
Under a scale transformation $x \to s \, x$ and 
$O_\U(sx) \to s^{-d_\U}O_\U(x)$, 
we have
\begin{eqnarray}
\label{normalization}
A_{d_\U} \  \theta(P^0) \ \theta(P^2) (P^2)^\alpha &=&   \int d^4 x s^4  e^{isP \cdot x} \langle 0| 
s^{-2d_\U}    O_\U(x)  O_\U^\dagger(0)|0\rangle \nonumber \\
& = & s^{-2(d_\U-2)} A_{d_\U} \ \theta(s P^0) \ \theta(s^2 P^2) \  (s^2 P^2)^\alpha  \; .
\end{eqnarray}
Requiring scale invariance implies $\alpha=d_\U-2 $, since the step 
functions are invariant.
Therefore, we obtain
\begin{equation}
\label{scalarO}
%\vert \langle 0 \vert O_\U (0)\vert P \rangle \vert^2
\rho_\U (P^2)  \, = \,
 A_{d_\U}\, \theta(P^0) \, \theta (P^2) \,(P^2)^{d_\U -2 }  \, \geq 0 \;,
\end{equation}
where $A_{d_\U}$ is normalized to interpolate 
the $d_\U$-body phase space of  massless
particle \cite{georgi1}.
The phase space factor for $n$ massless particle 
with
$(p_1 + p_2 + \cdots + p_n)^2 = s^2$ and
$p_i^2 = 0$
can be written as
\begin{equation}
{\mathrm dLIPS}_n=A_n s^{n-2} \ ,\quad
   A_n={16\pi^2\sqrt{\pi}\over (2\pi)^{2n}}
       { \Gamma(n+{1\over 2})\over\Gamma(n-1)\Gamma(2n)} \;,
\end{equation}
which for the first few $n$'s are 
$A_{n\to 1} \to 2\pi (n-1)$, $A_2={1\over 8\pi}$ and $A_3={1\over 256\pi^3}$, etc.
Based on the similar scale dependence, the unparticle spectral density
is identified with  the phase space of $d_{\cal U}$-body massless particle 
in a convention advocated in \cite{georgi1}:
$d_\U \to n$ and $A_n \to A_{d_\U}$.
So the factor $A_{d_\U}$ in Eq.~(\ref{normalization}) is given by
\begin{equation}
   A_{d_\U}={16\pi^2\sqrt{\pi}\over (2\pi)^{2{d_\U}}}
       { \Gamma({d_\U}+{1\over
       2})\over\Gamma({d_\U}-1)\Gamma(2\,{d_\U})}  \; .
\end{equation}
Note that $d_\U$ can now take on non-integral value as well.  
This is a peculiar feature of unparticle physics since one can now 
speak of something like fractional particles.

The differential cross section for a process involving the collision
of two massless particles in the initial state and producing an 
unparticle plus  a few other massless particles in the final state can 
be written as
 \[
 d \sigma(p_1,p_2 \to P_\U, k_1, k_2, ...) = \frac{1}{2s}\, 
|\overline{\cal M}|^2  \, d \Phi 
\]
where 
\begin{eqnarray}
d\Phi &=& (2\pi)^4 \, 
  \delta^{(4)}\left(p_1 + p_2 - P_\U - k_1 - k_2 - \cdots \right) \,
  \prod_{i} \left[ 2\pi \, \theta(k_i^0)\, \delta (k^2_i)\,
          \frac{d^4 k_i}{(2\pi)^4}             \right ]  \nonumber \\
 &   \times & \, A_{d_\U} \, \theta (P_\U^0) \, \theta (P_\U^2) \,
\left(P_\U^2\right)^{d_\U -2 } \,
         \frac{d^4 P_\U}{(2\pi)^4}
\label{5}
\end{eqnarray}
with $s = (p_1 + p_2)^2$ and $|\overline{\cal M}|^2$ is spin- and 
color-averaged matrix
element squared.  Note that in the limit $d_\U \to 1$ from above
\begin{equation}
  \lim_{d_\U \to 1^+}  A_{d_\U} \,(P^2_\U)^{d_\U-2} \, \theta(P_\U^0) \,
  \theta (P^2_\U)  = 
    2 \pi \theta(P_\U^0) \,\delta (P^2_\U) \;,
\end{equation}
so that the phase-space factor associated with the unparticle 
behaves just like a single massless particle in this limit.
If there are only one massless particle and an unparticle
in the final state, the phase space factor is further simplified to
\begin{equation}
 d\Phi = \frac{1}{2 (2\pi)^3} A_{d_\U} \, \theta (P_\U^0) \, 
 \theta (P_\U^2) \,
\left (P_\U^2 \right )^{d_\U -2} \, k_1^0 \, d k^0_1 d \Omega \;.
\end{equation}

\subsection{Virtual propagator of unparticle}

The derivation of the virtual unparticle propagator is also based on
scale invariance.  Without loss of generality we consider a 
scalar propagator.  The extensions to spin-1 and spin-2 propagators
simply include the appropriate spin structures and will be presented 
in the next subsection.
The Feynman propagator $\Delta_{F} (P^2)$ of the unparticle is determined 
by the spectral formula
\begin{eqnarray}
\label{lhz}
\Delta_{F} (P^2)&=&\frac{1}{2\pi} \, \int_0^\infty \, \frac{R(M^2) dM^2}
   {P^2-M^2 +i\epsilon}   \\
&=& \frac{1}{2\pi} \, - \!\!\!\!\!\! \int_0^\infty \frac{R(M^2) dM^2}{P^2-M^2} 
- i \frac{1}{2}\, R(P^2)   \theta(P^2) \;,
\label{lhz2}
\end{eqnarray}
where $R(M^2) = A_{d_\U} (M^2)^{d_\U -2}$ is the spectral density 
given in Eq.~(\ref{scalarO}).
The appropriate form for $\Delta_{F} (P^2)$ to be scale invariant 
is $\Delta_{F} (P^2) = Z_{d_\U} (-P^2)^{d_\U -2}$, 
where $Z_{d_\U}$ is the factor to be determined.
Note that our polar angle of complex number is restricted to $[-\pi, \pi)$. 
The complex function $(-P^2)^{d_\U -2}$ is analytic for negative $P^2$,
but needs a branch cut for positive $P^2$:
\begin{equation}
\label{branchcut}
 (-P^2)^{d_\U -2}=\left \{
\begin{array}{lcl}
|P^2|^{d_\U -2}   & \quad & \hbox{if $P^2$ is negative and real, } \\
|P^2|^{d_\U -2} e^{-i d_\U \pi} & & \hbox{for positive $P^2$ with 
an infinitesimal $i0^+$} . 
\end{array} \right.
\end{equation}
This choice guarantees a propagator with a  space-like momentum is real
without cuts.  We can then determine the factor $Z_{d_\U}$ by comparing 
with the
imaginary part of $\Delta_F (P^2)$ for a time-like momentum $(P^2>0)$:
\begin{equation}
\Im{\mathrm m} \Delta_F (P^2) = -Z_{d_\U} \sin(d_\U \pi) (P^2)^{d_\U -2} 
                  = - \frac{1}{2}A_{d_\U} \ (P^2)^{d_\U -2} \;.
\end{equation}
We thus obtain
\begin{equation}
  Z_{d_\U}  = \frac{A_{d_\U}} { 2 \sin (d_\U \pi) } \;,
\end{equation}
and the unparticle propagator is given by
\begin{equation}
\label{unpropagator}
 \Delta_F (P^2) =  \frac{A_{d_\U}} { 2 \sin (d_\U \pi) } (-P^2)^{d_\U -2} \;,
\end{equation}
where the definition of $(-P^2)^{d_\U -2}$ is given in Eq.~(\ref{branchcut}).
In $t$- or $u$- channel process, $(-P^2)$ is positive and 
so there is no complex phase associated with the propagator.  On the
other hand, for an $s$-channel process $(-P^2)$ is negative and so there
is a complex phase associated with the propagator.  This will lead to 
interesting interference effects with the standard model amplitudes.  
For instance, 
in $e^- e^+ \to \mu^- \mu^+$ \cite{georgi2} or Drell-Yan process \cite{ours},
the unparticle propagator can interfere with the real photon propagator 
and with 
both the real and imaginary parts of the unstable $Z$ boson propagator.
We note that since $Z_{d_\U} \to -1$ as $d_\U \to 1^+$, Eq.(\ref{unpropagator}) reproduces the familiar result
\begin{equation}
\label{scalarpropagator}
\lim_{d_\U \to 1^+}  \Delta_F (P^2)  = \frac{1}{P^2} \; .
\end{equation}

\subsection{Spin structures of unparticle operators}

In Eq.~(\ref{scalarO}), the operator $O_\U$ is a scalar.  It is 
straight-forward to extend to spin-1 and spin-2 unparticle operators by including appropriate 
tensor structures:
\begin{eqnarray}
%\langle 0 \vert O^\mu_\U (0)\vert P \rangle \, 
%\langle P \vert {O^\nu_\U}^\dagger (0)\vert 0 \rangle \, \rho(P^2)  
  \langle 0 \vert O^\mu_\U (x)  {O^\nu_\U}^\dagger (0) \vert 0 \rangle 
&=&
 A_{d_\U} \int \frac{ d^4 P}{(2\pi)^4}   \, e^{-i P \cdot x} \, 
  \theta(P^0) \, \theta (P^2) \,(P^2)^{d_\U -2 } \,
   \pi^{\mu\nu}(P) \;, \\
%%
%\langle 0 \vert O^{\mu\nu}_\U (0)\vert P \rangle \,
%\langle P \vert {O^{\rho\sigma}_\U}^\dagger (0)\vert 0 \rangle \, \rho(P^2)
  \langle 0 \vert O^{\mu\nu}_\U (x)  {O^{\rho\sigma}_\U}^\dagger (0) \vert 0 \rangle 
  &=& 
 A_{d_\U} \int \frac{ d^4 P}{(2\pi)^4}   \, e^{-i P \cdot x} \, 
  \theta(P^0) \, \theta (P^2) \,(P^2)^{d_\U -2 } \, 
               T^{\mu\nu,\rho\sigma}(P) \;,
\end{eqnarray}
where
\begin{eqnarray}
\label{spin1prop}
\pi^{\mu\nu}(P) &=& - g^{\mu \nu} + \frac{P^\mu P^\nu }{ P^2} \; ,  \\
T^{\mu\nu,\rho\sigma}(P) &=& \frac{1}{2} \, \left\{
   \pi^{\mu\rho}(P)\  \pi^{\nu\sigma}(P) 
        + \pi^{\mu\sigma}(P)\  \pi^{\nu\rho}(P) - \frac{2}{3}\ 
          \pi^{\mu\nu}(P)\  \pi^{\rho\sigma}(P)  \right\} \;.
          \label{spin2prop}
\end{eqnarray}
The forms of $\pi^{\mu\nu}(P)$ and $T^{\mu\nu,\rho\sigma}$  are chosen
such that $P_\mu \pi^{\mu\nu}(P) =0$,  $P_\mu T^{\mu\nu,\rho\sigma}(P) =0$, 
and $T^{\mu\; ,\rho\sigma}_{\;\;\mu} =0$. 
The unparticle operators are all taken to be Hermitian, and
$O^\mu_\U$ and $O^{\mu\nu}_\U$ are assumed to be transverse. 
In addition, the spin-2 unparticle operator is taken to be 
traceless $O^{\mu}_{\U \, \mu} = 0$.
The propagators for vector and tensor operators can be derived as in
Eq.~(\ref{unpropagator}) for the scalar case using spectral decomposition:
\begin{eqnarray}
 \left[ \Delta_F (P^2) \right]_{\mu\nu} &=&
         \frac{A_{d_\U}}{2 \sin (d_\U \pi)} \, (-P^2)^{d_\U -2} \, 
         \pi_{\mu\nu}(P) \;, \\
 \left[ \Delta_F (P^2) \right]_{\mu\nu, \rho\sigma} &=&
         \frac{A_{d_\U}}{2 \sin (d_\U \pi)} \, (-P^2)^{d_\U -2} \, 
         T_{\mu\nu, \rho\sigma}(P) \;.
\end{eqnarray}
%
%The spin-2 unparticle propagator has also been considered in 
%\cite{Mathews-Ravindran}.

\subsection{Effective operators}

\begin{figure}[t!]
\centering
\includegraphics[width=6in]{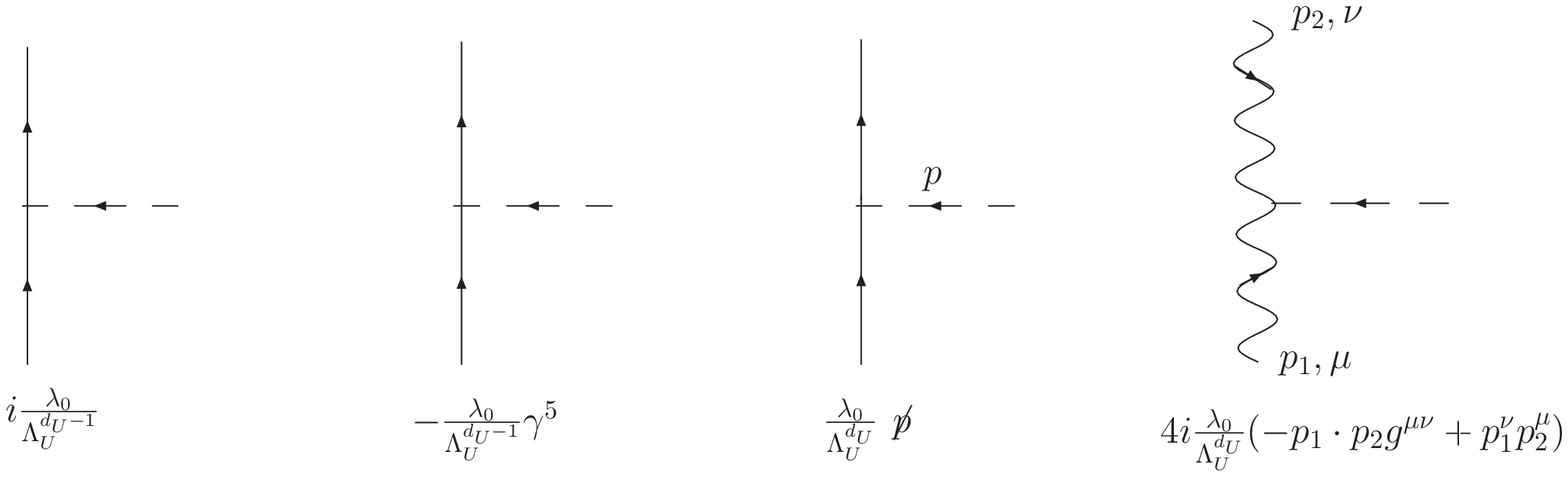}
\caption{\small \label{fey0}
Feynman rules for the scalar unparticle operators in Eq.~(\ref{lambda0}).}
\end{figure}

\begin{figure}[t!]
\centering
\includegraphics[width=4in]{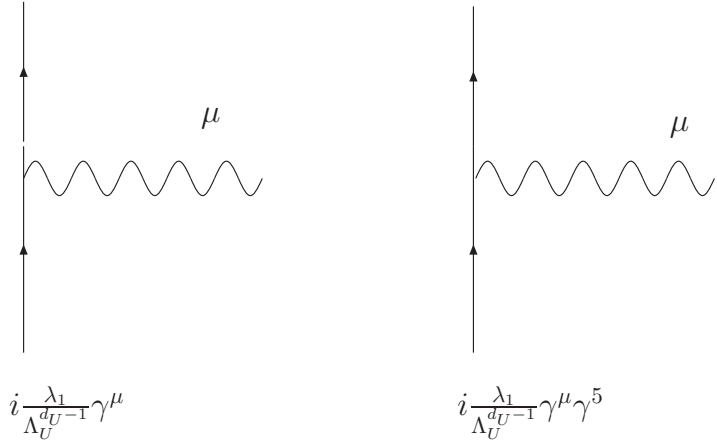}
\caption{\small \label{fey1}
Feynman rules for the vector unparticle operators in Eq.~(\ref{lambda1}).}
\end{figure}

\begin{figure}[t!]
\centering
\includegraphics[width=6in]{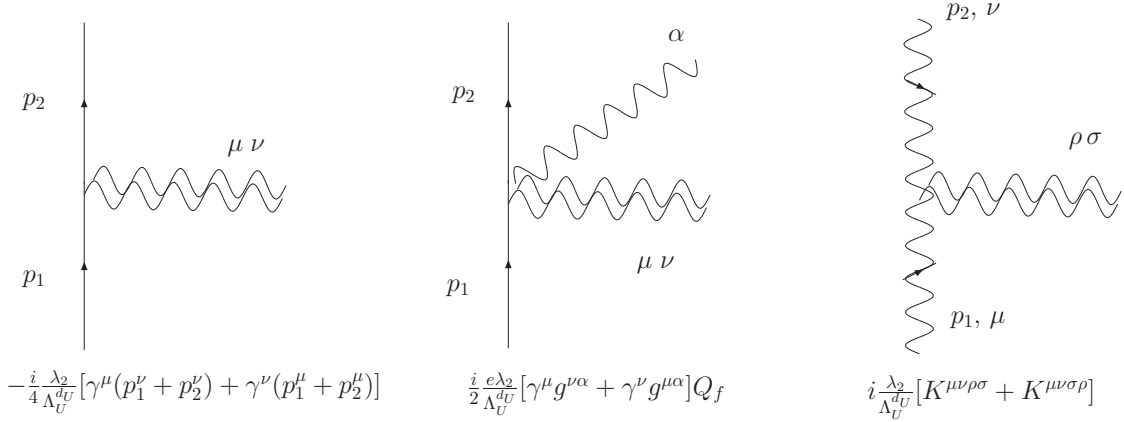}
\caption{\small \label{fey2}
Feynman rules for the tensor unparticle operators
in Eq.~(\ref{lambda2}).  The 
$K^{\mu\nu\rho\sigma} = -g^{\mu\nu} p_1^\rho p_2^\sigma - p_1 \cdot p_2 
g^{\rho\mu} g^{\sigma \nu} + p_1^\nu p_2^\rho g^{\sigma \mu}
                           + p_2^\mu p_1^\rho g^{\sigma \nu}$.
The double-wavy line represents a spin-2 unparticle while the single-wavy
line represents a photon, and  $Q_f$ denotes the electric charge of the 
fermion.  In case of a $Z$ boson in the middle diagram, replace $e Q_f$ by 
$\frac{g}{\cos\theta_{\rm w}}( T_{3f} - Q_f \sin^2\theta_{\rm w} )$,
where $T_{3f}$ is the isospin projection of the fermion doublet.
 }
\end{figure}

The common effective interactions 
that satisfy the standard model gauge symmetry 
for the scalar, vector,
and tensor unparticle operators with standard model fields are given, 
respectively, by
\begin{eqnarray}
&&
\lambda_0 \frac{ 1}{\Lambda_\U^{d_\U-1}} \bar f f O_\U\;, \;\;
\lambda_0 \frac{1}{\Lambda_\U^{d_\U-1} } \bar f i
      \gamma^5 f O_\U\;, \;\;
\lambda_0 \frac{1}{\Lambda_\U^{d_\U} } \bar f \gamma^\mu f 
(\partial_\mu O_\U) \;, \lambda_0 \frac{1}{\Lambda_\U^{d_\U} } G_{\alpha\beta} G^{\alpha\beta}
O_\U \;, 
\label{lambda0} \\
%%%%
%%%%
&&\lambda_1 \frac{1}{\Lambda_\U^{d_\U - 1} }\, \bar f \gamma_\mu f \,
O_\U^\mu \;, \;\;
\lambda_1 \frac{1}{\Lambda_\U^{d_\U - 1} }\, \bar f \gamma_\mu \gamma_5 f \,
O_\U^\mu \;, 
\label{lambda1} \\
%%%
%%%
&&- \frac{1}{4}\lambda_2 \frac{1}{\Lambda_\U^{d_\U} } \bar \psi \,i
   \left(  \gamma_\mu \stackrel{\leftrightarrow}{D}_\nu + 
           \gamma_\nu \stackrel{\leftrightarrow}{D}_\mu \right )
  \psi  \,  O_\U^{\mu\nu} \;,
   \lambda_2 \frac{1}{\Lambda_\U^{d_\U} } G_{\mu\alpha}
G_{\nu}^{\;\alpha} O_\U^{\mu\nu} 
 \;, 
\label{lambda2}
\end{eqnarray}
where the covariant derivative $D_\mu = \partial_\mu + i g \frac{\tau^a}{2} 
W^a_\mu + i g' \frac{Y}{2} B_\mu$, 
$G^{\alpha\beta}$ denotes the gauge field strength (gluon, photon and 
weak gauge bosons), $f$ stands
for a standard model fermion, $\psi$ stands for a standard model fermion doublet or singlet, 
and $\lambda_{i}$ are dimensionless effective
couplings $C_{O^i_\U} \Lambda_\U^{d_{\cal BZ}}/M_\U^{d_{\rm SM} + 
d_{\cal BZ}-4}$ with the index
$i=0,1$ and $2$ labeling the scalar, vector and tensor unparticle
operators, respectively.  Here we label each coupling constant 
$\lambda_{i}\; (i=0,1,2)$ the same for various operators of each spin.  
In principle, they can be different and they are then distinguished by additional indices.
For simplicity we will also assume universality
that $\lambda_i$'s are flavor blind. 
The Feynman rules for the operators in Eqs. (\ref{lambda0}), (\ref{lambda1}),
and (\ref{lambda2}) are shown in Figs. \ref{fey0}, \ref{fey1}, 
and \ref{fey2}, respectively.
Conventional wisdom tells us that the scalar operator $O_\U$ coupled
 to fermion is suppressed by the fermion mass.
As already studied in \cite{georgi1},\cite{georgi2},\cite{ours}, some of the
operators listed above can give rise to interesting phenomenology,
including real emission of unparticle as well as effective 4-fermion
contact interactions.  Phenomenology of unparticles that couple to
flavor changing neutral currents have also been studied in
\cite{georgi1},
\cite{Luo-Zhu},
\cite{Chen-Geng},
\cite{Aliev-Cornell-Gaur},
\cite{Li-Wei},
\cite{Lu-Wang-Wang},
\cite{Choudhury-Ghosh-Mamta}. More gauge invariant operators that couple the 
spin-0 and spin-1 unparticle operators
to SM fields are listed in \cite{Chen-He}.

\subsection{Effective four-fermion interactions}

\begin{figure}[t!]
\centering
\includegraphics[width=4.75in]{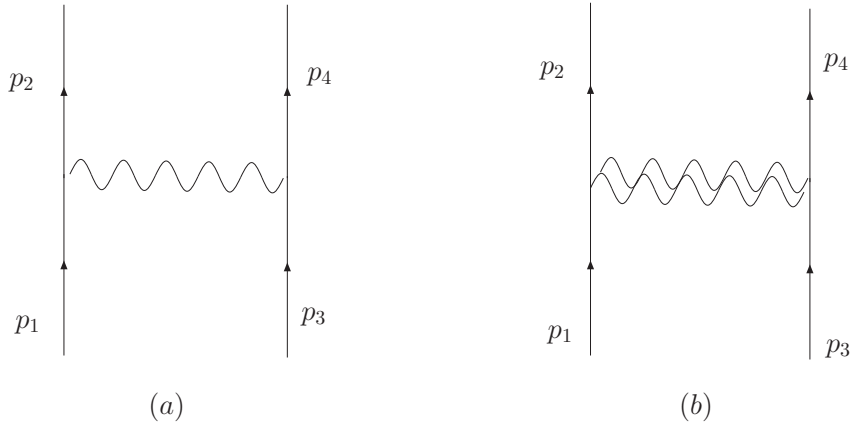}
\caption{\small \label{4fermion}
Feynman diagrams for exchange of a spin 1 and spin-2 unparticles between two fermionic
currents.}
\end{figure}

Virtual exchange of unparticle corresponding to the vector operator
$O_\U^\mu$ between two fermionic currents
can result in the following 4-fermion interaction (Fig. \ref{4fermion}a) \cite{ours}
\begin{eqnarray}
{\cal M}_1^{4f} = \lambda_1^2 \, Z_{d_\U} 
\, \frac{1}{\Lambda_\U^2} \,
 \left(
- \frac{P_\U^2}
{\Lambda_\U^2} 
  \right)^{d_\U - 2} \, 
( \bar f_2 \gamma_\mu f_1)\,  ( \bar f_4 \gamma^\mu f_3) \; .
\label{4fermionsop}
\end{eqnarray}
The 4-momentum flowing along the unparticle propagator is $P_\U \equiv (p_1 -
p_2)$.  
The contribution from the longitudinal piece 
$P_\U^\mu P_\U^\nu/P_\U^2$ 
in Eq.~(\ref{spin1prop})
has been dropped for massless external fermions.
The convention of the fractional exponent of a complex number is already 
given in Eq.~(\ref{branchcut}).
The $(-)$ sign in front of $P_\U^2$ of 
the unparticle propagator in  Eq.~(\ref{4fermionsop})
gives rise to a phase factor $\exp (-i\pi d_{\cal U})$ for  time-like
momentum  $P_{\cal U}^2>0$, 
but not for   space-like momentum $P_{\cal U}^2<0$.  
For example, in Drell-Yan production the virtual exchange of unparticle
in the $s$-channel will have $P_\U^2$ taken as the $\hat s$ of the
subprocess and therefore will contain a phase.
The most important feature is that the high
energy
behavior of the amplitude scales as $(\hat s/\Lambda_\U^2 )^{d_\U  -1}$.  
For $d_\U=1$ the tree amplitude behaves like that of a massless
photon exchange, 
while for $d_\U = 2$ the amplitude reduces to the
conventional 4-fermion interaction \cite{Eichten},\cite{contact}, 
i.e., its high-energy behavior
scales like
$s/\Lambda_\U^2$.  If $d_\U$ is between 1 and 2, say $3/2$, the
amplitude has the unusual behavior of $\sqrt{\hat s}/\Lambda_\U$ at
high energy.  
If $d_\U = 3$ the amplitude's high energy behavior becomes $(\hat
s/\Lambda_\U^2)^2$,
which resembles the exchange of Kaluza-Klein tower of gravitons
\cite{km}.
In principle, we can allow different
couplings in different chirality combinations in the 4-fermion contact
interactions, denoted by $LL,RR,LR,RL$, which can produce
parity violation and therefore the forward-backward asymmetries.
The combination
of $LL+RR+LR+RL$ gives $VV$ interaction while $LL+RR-LR-RL$ gives
$AA$ interaction that correspond to the vector and axial-vector interactions
introduced in Ref. \cite{georgi2}. 

One can also consider the exchange of spin-2 unparticle between a pair of
fermionic currents.  The operator is given in Eq.~(\ref{lambda2}) and 
the Feynman rule in Fig. \ref{fey2}.  After simplification we arrive at
the following 4-fermion interaction
\begin{eqnarray}
{\cal M}_2^{4f} &=& - \frac{1}{8}\,\lambda_2^2 \, Z_{d_\U} 
\, \frac{1}{\Lambda_\U^4} \, 
 \left(
- \frac{ P_\U^2}{\Lambda_\U^2} \right)^{d_\U - 2} 
\, \left( \bar f_2 \gamma^\mu f_1 \right)\,\left( \bar f_4 \gamma^\nu f_3
   \right)\,
\nonumber \\
&\times&
 \left[(p_1 + p_2 ) \cdot (p_3 + p_4) g_{\mu\nu} + (p_1+p_2)_\nu (p_3+p_4)_\mu
 \right ] \;, \label{4-fermion-spin2}
\end{eqnarray}
for massless external fermions, where $p_i$ denotes the 4-momentum of
the fermion $f_i$ along the fermion line (Fig. \ref{4fermion}b). 
Note that the 4-fermion interaction induced by the spin-2
unparticle operator is further suppressed by $(s/\Lambda_\U)^2$
relative to that induced by spin-1 unparticle operator.  This is
similar to the exchange by a spin-2 graviton (which corresponds
exactly when $d_\U$ is set to 2 in Eq.~(\ref{4-fermion-spin2}).)  
Similarly, different chirality combinations are
possible for the 4-fermion contact interactions with spin-2 unparticle
exchange. 

The above 4-fermion amplitudes can interfere
with the standard model amplitudes of $\gamma$, $W$ and $Z$ exchange, and
thus leads to interesting interference effects.  In particular, the different
spin structures could be differentiated by studying various angular
distributions. Based on these spin-1 and spin-2 unparticle exchange amplitudes one 
can study the Drell-Yan process
at hadron colliders, deep-inelastic scattering at $ep$ colliders,
fermion pair production at $e^- e^+$ colliders, atomic parity violation, 
as well as many other low-energy $eq$ scattering processes, just in similar
ways as the conventional 4-fermion contact interactions \cite{contact}
or as the Kaluza-Klein states of graviton \cite{km2}.
Modification of the Newton's inverse square law 
in the sub-millimeter range due to spin-2 unparticle exchange 
and its possible tests at low energy gravity experiments
have  been studied in \cite{ungravity}.

\subsection{Conjecture to large extra dimensions}

The close similarity between 
the unparticle and 
Kaluza-Klein (KK) modes of the large extra
dimensions \cite{Arkani-Hamed:1998rs} (LED)
has been recognized \cite{ours}
in the calculation of the production cross sections and 
in virtual effects.
%%%%%
The unparticle and the KK states \cite{Han:1998sg},\cite{Giudice:1998ck} share
analogous phase space integrations \cite{km}, in particular
the integration over the invariant mass squared $P^2$.  It would be
interesting to relate the unparticle with the KK modes in LED.

Let us first set up  all fields of the standard model to be confined
on a flat 3 dimensional spatial brane with coordinates $\bf x$. 
A scalar unparticle field can be identified as 
a massless scalar bulk field $\Phi(t,{\bf x},y)$ permeating  
into the LED described by extra coordinates $y_i$ $(i=1,\cdots, n)$.
We study the simplest case that the space of LED is flat and periodic
in each $y_i$ with periodicity  $L$. The massless energy-momentum relation is
\begin{equation} 
E^2={\bf p}^2+ \sum_{i=1}^n (k_i)^2 \ , 
\end{equation}
where ${\bf p}$ is the momentum in the ordinary 3-space and $k_i$
is the momentum component in LED. 
Periodic conditions on the extra dimensions require all the 
momenta $k_i$ to be quantized such that they are integral multiples of  
$2\pi  /L$.
As SM physics only operates on  the 3-brane, the term 
$\sum_{i=1}^n (k_i)^2$ of the corresponding KK modes effectively becomes
the mass-squared of a particle propagating in the 3+1 spacetime.
For large $L$, the summation over the KK modes turns into an integral 
and the density of states is introduced as
\begin{equation} 
\sum_{\vec k} \longrightarrow
\int \left( \frac{L}{2\pi} \right)^n  d^n k   
\; = \;    
\int {L^n (m^2)^{{n\over2}-1} dm^2 \over (4\pi)^{n\over2} \Gamma(\frac{n}{2})} \; . 
\end{equation}
Identifying the power of $m^2$ in the density of states with 
the power of $P^2$ in the spectral density of the unparticle, we obtain
\begin{equation}
\label{yeeconjecture}
d_{\cal U}  =\frac{n}{2}+1  \; .
\end{equation}  
%
%Therefore, we have $d_{\cal U}={3\over2}$ or  2 for $n=1$ or 2
%respectively, for example. 
%
With one extra dimension we can have the notion 
of one-and-a-half particle viewed from the 3-brane, and so on.
It is also tempting to make the following identification
\begin{equation}
\label{ledunparticle}
A_{d_\U} = {L^{2 (d_\U - 1)} \over (4\pi)^{d_\U - 1} \Gamma(d_\U - 1)}
\end{equation}
with $d_\U$ given by Eq.(\ref{yeeconjecture}). 
Perhaps hidden higher dimension spacetime reveals itself through 
the unparticle physics.
It might be interesting to see if realistic models can be built based on this 
alternative interpretation of unparticle.
%Phenomenology of unparticle physics defined by 
%Eqs.(\ref{yeeconjecture}) and (\ref{ledunparticle}) 
%might be interesting too.
Recently, it has been demonstrated in \cite{Stephanov} that 
other values of $d_{\cal U}$ related to a dimensionless mass parameter
can be achieved by deconstructing 
the unparticle in the 5 dimensional warped anti-de Sitter space 
using the AdS/CFT correspondence.

\section{Phenomenology: Real Emission}

\subsection{Mono-photon and mono-$Z$ production in $e^- e^+$ collisions}

The energy spectrum of the mono-photon from the process 
$e^- (p_1) \; e^+(p_2)\; \to \gamma
(k_1) \; \U (P_\U)$ can be used to probe the unparticle \cite{ours}.
Similarly, the mono-$Z$ production is also sensitive to the presence
of some unknown particles or unparticle.  Let us first  derive the cross  
section
formulas for mono-$Z$ production.

The differential cross section for $f (p) \, \bar f(p')  \to Z(k) \, \U(P_\U)$
is given by
\begin{equation} 
\label{ffZUxsection}
 d \sigma = \frac{1}{2 s} \,  |\overline{ {\cal M}} |^2 \;
 \frac{ \sqrt{E_Z^2 - M_Z^2} A_{d_\U} }{ 16 \pi^3 \Lambda_\U^2} 
 \left( {P^2_\U\over \Lambda_{\cal U}^2} \right )^{ d_\U - 2}  
\,  \theta (P_{\U}^0) \theta (P^2_\U) d E_Z d \Omega_Z  \;,
\end{equation}
where $|\overline{ {\cal M}} |^2$ is the spin- and color-averaged
matrix element squared.
Note that the invariant mass squared $P^2_\U$ 
of the unparticle is not fixed but is 
related to the energy $E_Z$ of the $Z$ boson via the
recoil mass relation, 
\begin{equation}  P^2_\U = s +M_Z^2 - 2 \sqrt{s} \, E_Z  \; ,
\label{kinematic}
\end{equation}
where the energy range of $E_Z$ is 
\begin{equation}
M_Z \leq E_Z \leq E_Z^{\rm max} \equiv \frac{s+M_Z^2}{2\sqrt s} \; .
\end{equation}
As usual, we define $s=(p+p')^2$, $t=(p-k)^2$ and $u=(p-P_\U)^2$. 
Moreover, $s+t+u = M_Z^2 + P_\U^2$.

As $d_\U$ approaches unity, we recover the on-mass-shell condition in the
phase space
\begin{equation}
\lim_{d_\U \to 1^+}  A_{d_\U} (P^2_\U)^{d_\U-2} \theta (P^2_\U) = 
2 \pi \delta (P^2_\U) 
 =  \frac{1}{2 \sqrt s} \delta (E_Z - E_Z^{\rm max}) \;.
\end{equation}
Thus, the integral over $E_Z$  is trivial and the cross section becomes
\begin{equation} 
 \lim_{d_\U \to 1^+} d \sigma = \frac{1}{2 s} \,
 \frac{1}{32 \pi^2}  \left( 1 - \frac{M_Z^2}{s} \right) \,
   | \overline{{\cal M}} |_{E_Z = E_Z^{\rm max}}^2 \;
  d \Omega_Z  \;.
\end{equation}
This reproduces the usual formula for $2\to 2$ cross
section. This is expected since $d_\U \to 1$ corresponds to unparticle
$\to$ particle. In this case, the energy spectrum for
the $Z$ boson is just a delta function localized at $E_Z = E_Z^{\rm max}$.

\subsubsection{Spin-1 unparticle}

Let us turn our focus back to unparticle.  For spin-1 unparticle, we
consider only the first (vectorial) operator in Eq.~(\ref{lambda1}). Including the
second (axial-vectorial) operator in Eq.~(\ref{lambda1}) is straightforward.  There are
two contributing Feynman diagrams, $t$- and $u$-channels.  The matrix
element squared for $f (p) \bar f(p') \to Z(k) \U(P_\U)$ is given by
\begin{eqnarray}
  |\overline{{\cal M}}|^2 & = &
  \frac{2}{N_c}  \lambda_1^2 
  \frac{ e^2 ( {g_{L}^f}^2 + {g_{R}^f}^2) }
 { \sin^2 \theta_{\rm w} \cos^2 \theta_{\rm w}}   
  g\left( t / M_Z^2, u / M_Z^2, P^2_\U / M_Z^2 \right)
 \end{eqnarray}
where $N_c$ is the number of color for the fermion $f$, 
$g^{f}_L = T_{3f} - Q_f \sin^2 \theta_{\rm w}$, $g^f_R = - Q_f 
\sin^2 \theta_{\rm w}$ with $Q_f$ is the electric charge of the 
fermion $f$ and the function $g(x,y,z)$ is defined by 
\begin{equation}
g(x,y,z) = \frac{1}{2} \left( \frac{x}{y} + \frac{y}{x} \right) + 
  \frac{(1 + z)^2}{xy}
-\frac{z}{2} \left(  \frac{1}{x^2} + \frac{1}{y^2} \right) - 
\left(1 + z \right) \left(\frac{1}{x} + \frac{1}{y} \right) \; .
\label{flemingfun}
\end{equation}

The result for $f (p) \bar f(p')  \to \gamma(k)  \U(P_\U)$ can be
obtained by setting $M_Z$ equal to zero and appropriate substitution for
the couplings in Eq.~(\ref{ffZUxsection}), viz.
\begin{equation} 
 d \sigma = \frac{1}{2 s} \,  | \overline{{\cal M}} |^2 \;
 \frac{ A_{d_\U} }{ 16 \pi^3 \Lambda_\U^2} 
 \left( {P^2_\U\over \Lambda_{\cal U}^2} \right )^{ d_\U - 2}  
\, E_\gamma d E_\gamma d \Omega  
\end{equation}
with the matrix element squared given by
\begin{equation}
  |\overline{{\cal M}}|^2 = \frac{2}{N_c} \lambda_1^2 e^2 Q_f^2  \,
  \frac{ u^2 + t^2 + 2 s P^2_\U}{ u t} \; .
  \end{equation}
The $P^2_\U$ is related to the energy of the photon $E_\gamma$ by a simpler
recoil mass relation,
\begin{equation}  P^2_\U = s - 2 \sqrt{s} \, E_\gamma \ . 
\end{equation}

\begin{figure}[t!]
\centering
\includegraphics[width=3.2in]{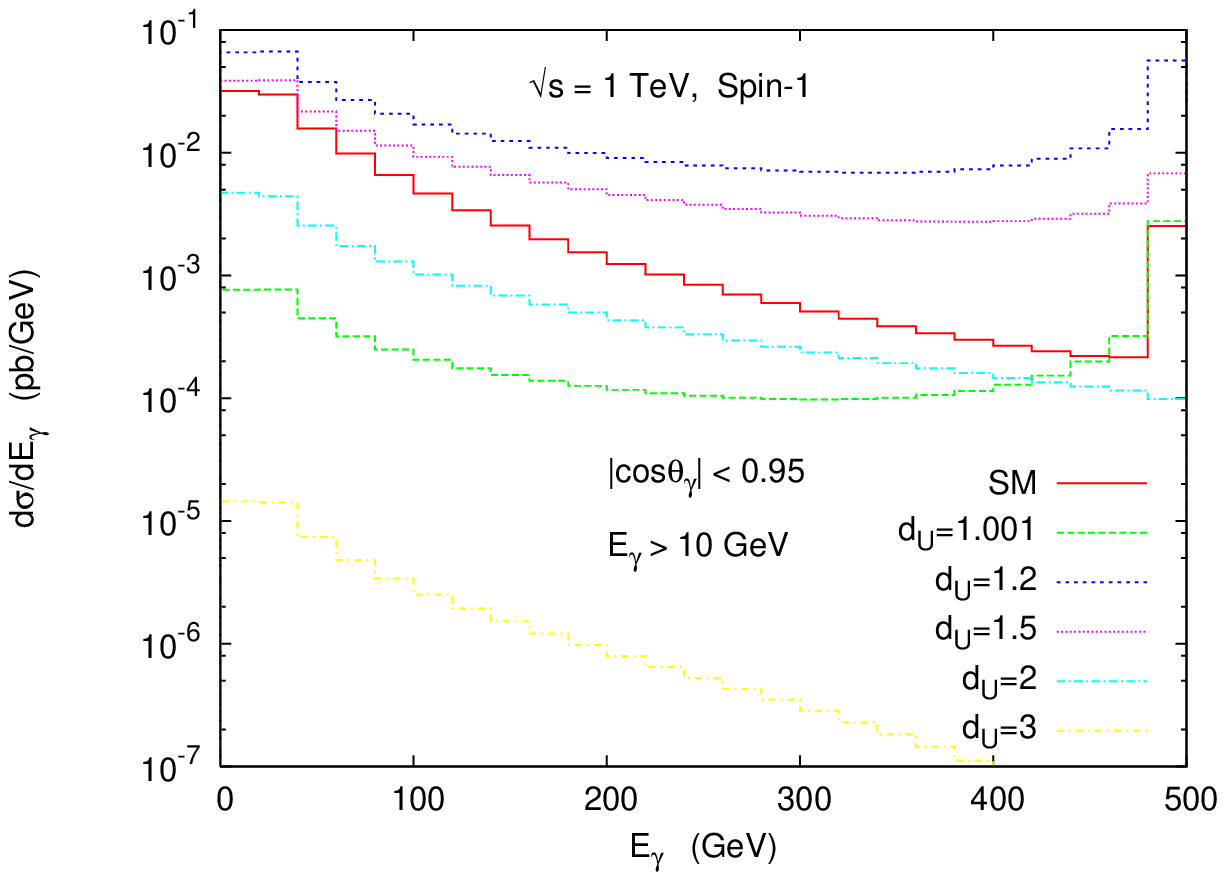}
\includegraphics[width=3.2in]{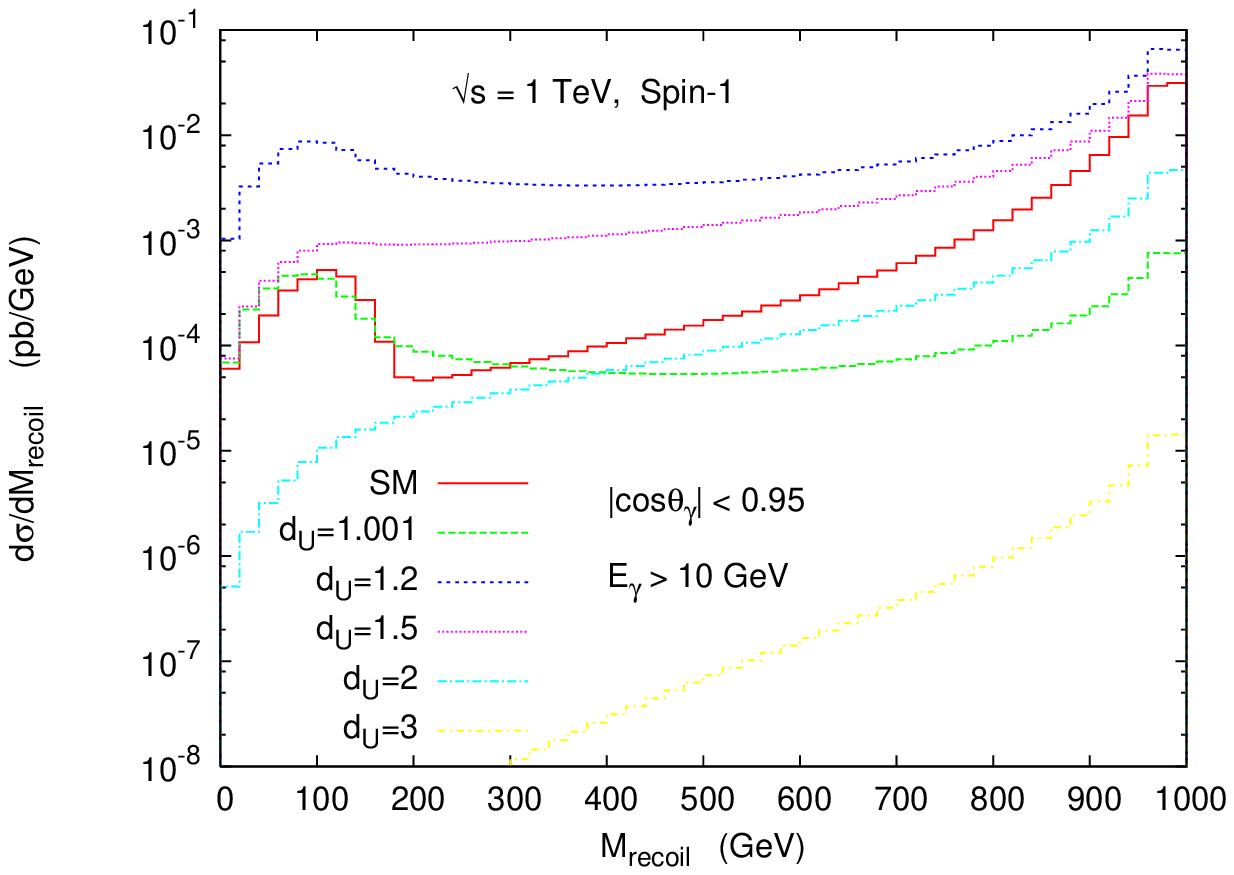}
\caption{
\label{eegammaU-spin1} \small
Comparison of photon energy and recoil mass distributions 
of $e^- e^+ \to \gamma {\cal U}$ (spin-1 unparticle) with 
the standard model background $e^-e^+\to \gamma Z^* \to \gamma \nu \bar \nu$ 
for different values of $d_{\cal U}=1.001,\,1.2,\,1.5,\,2$ and 3 at 
$\sqrt s = 1$ TeV.
}
\end{figure}

The mono-photon energy and recoil mass distributions are
plotted in Fig.~\ref{eegammaU-spin1} for various choices of $d_\U$ 
at $\sqrt s = 1$ TeV. The
sensitivity of the scale dimension to these distributions can be
easily discerned.  The standard model background from 
$e^- e^+ \to \gamma Z^* \to \gamma \nu \bar \nu$ is also displayed 
for comparison.
Similar features are also found for the process $e^-e^+ \to Z \U$ which 
has also been studied recently in \cite{Chen-He}.

\subsubsection{ Spin-2 unparticle}

We consider both spin-2 unparticle operators in Eq.~(\ref{lambda2}) and 
let their coupling constants be different, denoted by
$\lambda'_2$ and $\lambda_2$, respectively.  
There are four contributing Feynman 
diagrams for the process: $t$- and $u$-channels plus a seagull diagrams 
from the first operator
and an $s$-channel diagram from the second.
The matrix element squared for 
$f (p) \, \bar f(p')  \to Z(k) \, \U(P_\U)$ is given by
\begin{eqnarray}
  |\overline{{\cal M}}|^2 & = &
  \frac{1}{4N_c}   \frac{ \lambda_2^2}{\Lambda^2_\U}
  \frac{ e^2 ( {g^f_L}^2 + {g^f_R}^2) }
               { 2 \sin^2 \theta_{\rm w} \cos^2 \theta_{\rm w}}   
  \frac{1}{3 (s-M_Z^2)^2 t^2 u^2}
  \left[ 
  F(t,u) + r G (t,u) + r^2 H(t,u)
  \right]
 \end{eqnarray}
with $r = \lambda^\prime_2 / \lambda_2$ and
\begin{equation}
(F,G,H) =(F_0,G_0,H_0)+
\frac{1}{P^2_\U}(F_2,G_2,H_2) +\frac{1}{P^4_\U}(F_4,G_4,H_4) \;,
\end{equation}
where these complicated functions can be found in the appendix.
We note that these functions satisfy the following equations
\begin{eqnarray}
F_2 + G_2 + H_2 & = & 0 \; , \nonumber \\
F_4 + G_4 + H_4 & = & 0  \; .\nonumber
\end{eqnarray}
Thus, if we set $r=1$, i.e., $\lambda_2=\lambda'_2$, 
the $1/P^2_\U$ and $1/P^4_\U$ terms in the matrix element squared 
summed up to zero. 
This reflects the fact that 
the longitudinal parts in polarization sum of the spin-2 unparticle are 
just like the gauge artifact of the spin-2 massless graviton. 
They should not contribute to physical matrix elements.
Note that the longitudinal part of the polarization sum of the $Z$ boson 
does not contribute to the matrix element squared 
either, because the external fermion masses are set to be zero.
In the case of $r=1$, the above matrix element squared is simplified to
\begin{eqnarray}
  |\overline{{\cal M}}|^2 & = &
  \frac{1}{4N_c}   \frac{ \lambda_2^2}{\Lambda^2_\U}
  \frac{ e^2 ({g^f_L}^2 + {g^f_R}^2) }
  { 2 \sin^2 \theta_{\rm w} \cos^2 \theta_{\rm w}}   
  \frac{1}{3 (s-M_Z^2)^2 t^2 u^2} {\cal F} (t,u)
  \label{matrixelmKKa}
 \end{eqnarray}
where
\begin{eqnarray}
{\cal F} & \equiv & F_0 + G_0 + H_0 \nonumber \\
& = & 
8 M_Z^6 t u  \left[ 3 P_\U^4 + 4 t u -  3 P_\U^2 \left( t + u \right) \right] \nonumber \\
 & + & 3 t u \left( -P_\U^2 + t + u \right) 
 \left[ 2 P_\U^4 + t^2 + u^2 - 2 P_\U^2 \left(t + u\right) \right]
 \left[ -P_\U^4 - 4 t u + P_\U^2 \left(t +  u \right) \right] \nonumber \\
         & + & 2 M_Z^4 t u \left[ 27 P_\U^6 - 42 P_\U^4 \left( t + u \right) - 28 t u \left( t + u \right) 
         + 5 P_\U^2 \left( 3 t^2 + 16 t u + 3 u^2 \right) \right] 
        \nonumber \\ 
        &+ & M_Z^2 \left[ 52 t^3 u^3 + 
          36 t^2 u^2 \left( t^2 + u^2 \right) - 3 P_\U^8 \left( t^2 - 12 t u + u^2 \right)  \right.
             \nonumber \\
          &-& 6 P_\U^2 t u \left( t^3 + 23 t^2 u + 23 t u^2 + u^3 \right)
          -  3\ P_\U^4 \left( t^4 - 14 t^3 u - 62 t^2 u^2 - 14 t u^3 +  u^4\right) 
                        \nonumber \\
           &+&  \left. 6 P_\U^6 \left( t^3 + u^3 - 12 t u \left( t + u \right) \right) \right]  \; .
  \label{matrixelmKKb}
\end{eqnarray}
Equations (\ref{matrixelmKKa})--(\ref{matrixelmKKb}) coincide with the 
matrix element for $f \bar f \to Z G$ where $G$ is the Kaluza-Klein
graviton obtained previously in \cite{CK}. Setting $r=1$ implies that the 
two operators in Eq.(\ref{lambda2}) sum up and has the form of the 
energy-momentum stress tensor in flat spacetime. This idea has been generalized to 
curved spacetime \cite{ungravity}.

Just like the spin-1 case, we can obtain 
$f (p) \bar f(p')  \to \gamma(k) \, \U(P_\U)$ with appropriate substitutions:
\begin{eqnarray}
  |\overline{{\cal M}}|^2 & = &
  \frac{1}{4N_c}   \frac{ \lambda_2^2}{\Lambda^2_\U}
  e^2 Q_f^2 
  \frac{1}{3 s^2 t^2 u^2}
  \left[ 
  F(t,u) + r G (t,u) + r^2 H(t,u)
  \right] \;,
 \end{eqnarray}
where $F, G$ and $H$ are given by the previous formulas with $M_Z$ setting
 to zero. 
In the case of $r=1$, the matrix element squared reduces to
\begin{eqnarray}
  |\overline{{\cal M}}|^2 & = &
  \frac{1}{4N_c}  \frac{ \lambda_2^2}{\Lambda^2_\U}
  e^2 Q_f^2 
  \frac{1}{s t u} 
   \left( 2 \, s \, P_\U^2 + t^2 + u^2  \right)
 \left( s \, P_\U^2 + 4 \,t \, u  \right)  \;.
 \end{eqnarray}

\begin{figure}[t!]
\centering
\includegraphics[width=3.2in]{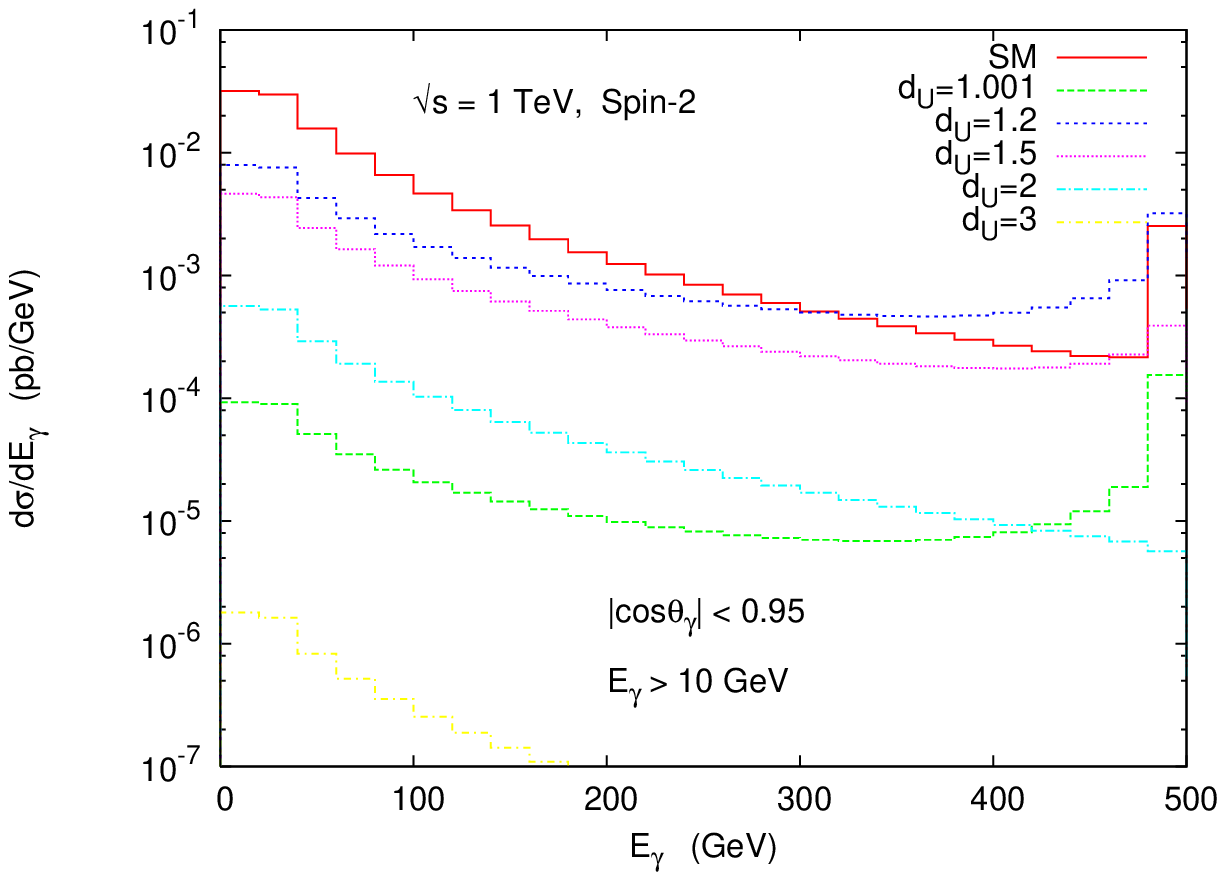}
\includegraphics[width=3.2in]{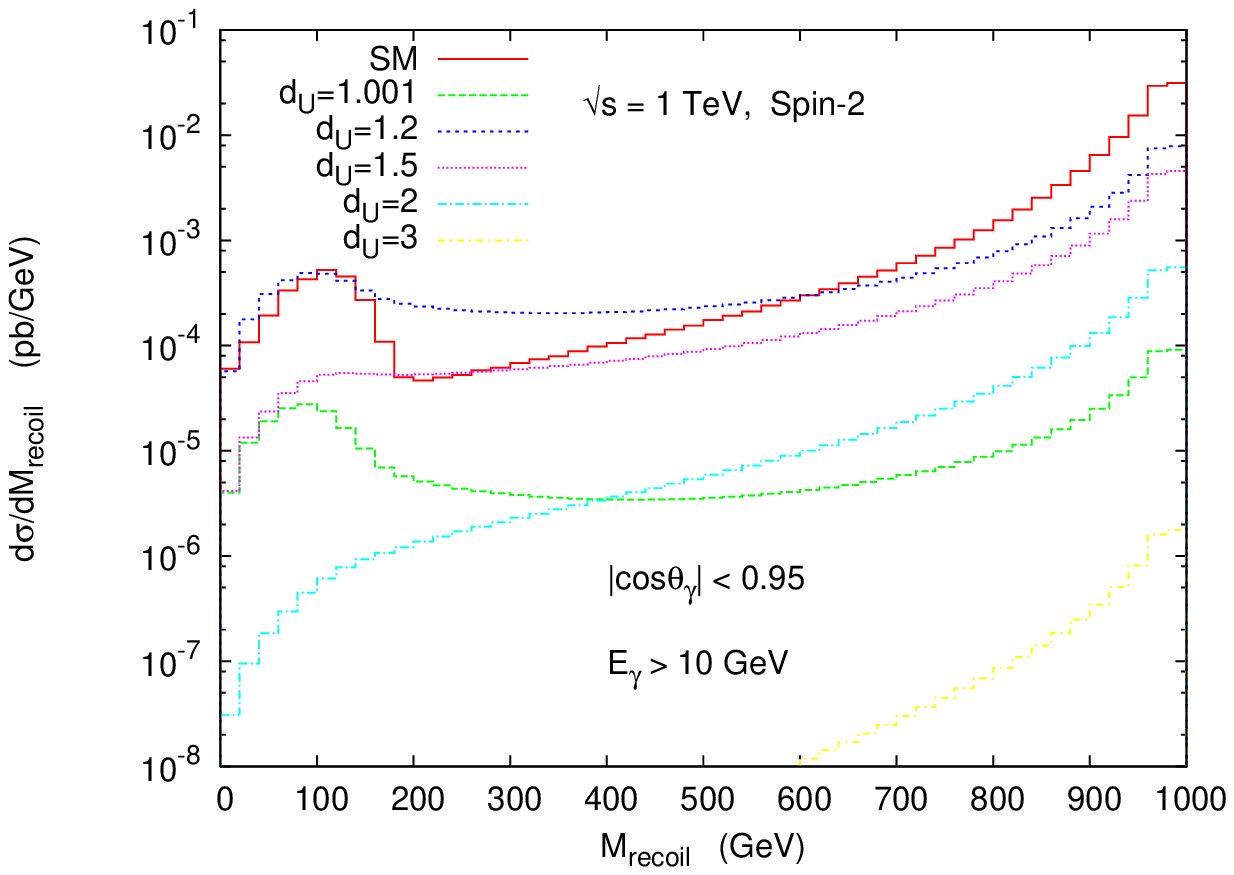}
\caption{
\label{eegammaU-spin2} \small
Comparison of photon energy and recoil mass distributions of $e^- e^+ \to \gamma {\cal U}$ (spin-2 unparticle) with the
standard model background $e^-e^+\to \gamma Z^* \to \gamma \nu \bar \nu$ 
for different values of $d_{\cal U}=1.001,\,1.2,\,1.5,\,2$ and 3 at $\sqrt s = 1$ TeV.
}
\end{figure}

The mono-photon energy and recoil mass distributions for emission of spin-2
unparticle are
plotted in Fig.~\ref{eegammaU-spin2} for various choices of $d_\U$ at
$\sqrt s =$ 1 TeV with $r = 1$. 
The sensitivity of the scale dimension to these distributions can be also
easily discerned.  The standard model background from 
$e^- e^+ \to \gamma Z^* \to \gamma \nu \bar \nu$ is also displayed 
for comparison. 
Similar features are also found for the process $e^-e^+ \to Z \U$ for the 
spin-2 case.

\subsection{$Z \to f \bar f \U$} 
\begin{figure}[t!]
\includegraphics[width=4.5in]{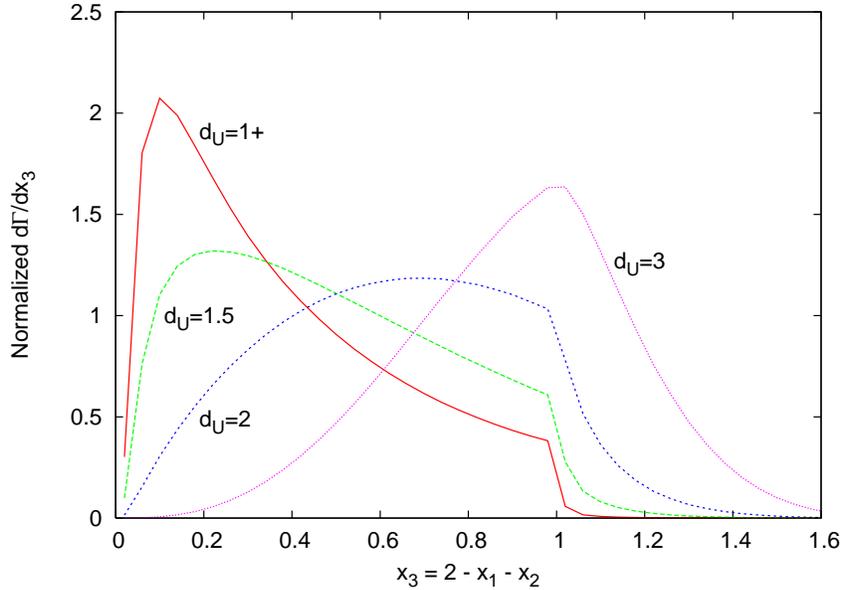}
\caption{
\label{zqqu} \small
Normalized decay rate of $Z \to q \bar q {\cal U}$  for spin-1 unparticle 
versus $x_3 = 2 - x_1 - x_2$ for different values of 
$d_{\cal U}=1^+,\,1.5,\,2$, and 3, where ``$1^+$'' stands for $1+\epsilon$
for a small positive $\epsilon$.
}
\end{figure}

The decay width for the process $Z \to f \bar f \U$ with a spin-1 
unparticle can be
easily obtained as \cite{ours}
\begin{eqnarray}
\label{ztoqqU}
\frac{d\Gamma(Z \to f \bar f +{\cal U}) }
{ d x_1 d x_2 d \xi} & =
 &
 \Gamma( Z \to f \bar f) \frac{\lambda_1^2}{8\pi^3} \, g(1-x_1,1-x_2,\xi)  \frac{M_Z^2} {\Lambda_\U^2}
 A_{d_{\cal U}} \left(\frac{P_\U^2}{\Lambda_\U^2}\right)^{d_{\cal
 U}-2}
\end{eqnarray}
where $ \xi = P_\U^2/M_Z^2$ and $x_{1,2}$ are the energy fractions of
the fermions $x_{1,2} = 2 E_{f, \bar f} / M_Z$. The function
$g(z,w,\xi)$ has been defined in Eq.(\ref{flemingfun}).
The integration domain for Eq.~(\ref{ztoqqU}) is defined by $0 < \xi <
1, 0 < x_1 < 1 - \xi$ and $1- x_1 - \xi < x_2 < (1 - x_1 - \xi )/( 1 -
x_1)$.
In \cite{ours}, we plotted  the normalized decay rate of this
process versus the energy fraction of the fermion $x_1$.
Here, in Fig.~\ref{zqqu}, we plot the normalized decay rate of this
process versus the energy fraction of the unparticle $x_3 = 2 - x_1 - x_2$. 
One can see that the shape depends sensitively on the scale dimension of the
unparticle operator. As $d_\U \to 1$, the result approaches to a
familiar case of $\gamma^* \to q \bar q g^*$ \cite{cky}.  
The matrix element squared for $Z \to f \bar f \U$ with spin-2 unparticle can be obtained by 
applying crossing symmetry to the formulas for  $f \bar f  \to Z \U$ given 
in Eqs.(\ref{matrixelmKKa})--(\ref{matrixelmKKb}). 
We omit the detailed formulas here.

\subsection{Mono-jet production at hadronic collisions} 
\begin{figure}[t!]
\includegraphics[width=4.5in]{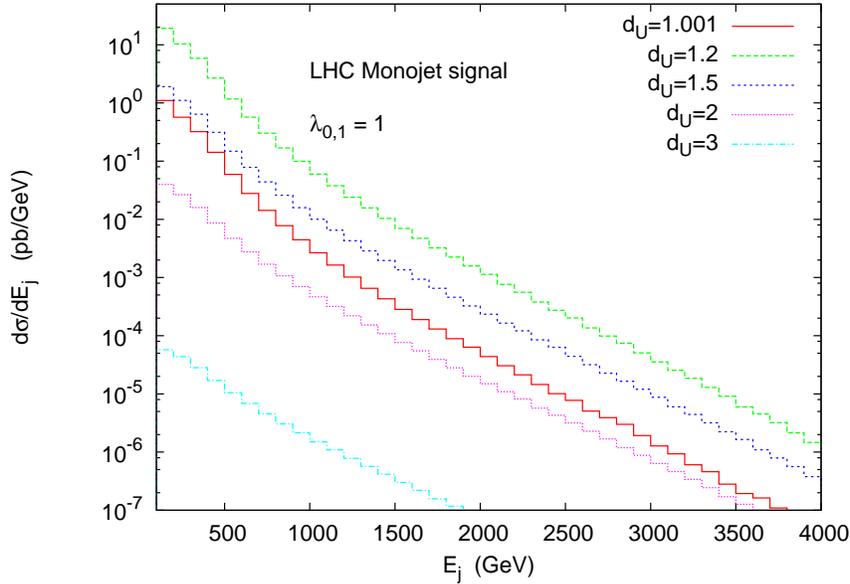}
\caption{
\label{monojet} \small
Differential cross section $d\sigma/dE_j$ versus $E_j$ for the monojet signal
at the LHC, with various $d_\U$.  
We have set $\Lambda_\U = 1$ TeV and $\lambda_0 = \lambda_1 = 1$.
}
\end{figure}

It was suggested in
\cite{georgi1} that in hadronic collisions the following partonic
subprocesses which can lead to mono-jet signals could be important for
detection of the unparticle.
\begin{eqnarray}
 g      g  & \to &  g \U   \;,\; \;    q \bar q   \to  g \U     \;
 ,\nonumber \\ 
 q      g  & \to & q \U  \;,\; \; \bar q g \, \to \, \bar q \U
 \;. \nonumber
\end{eqnarray}
For the subprocesses that involve both quark and gluon,
we consider solely the effects from the vector operator
$O^\mu_\U$. 
For the gluon-gluon fusion subprocess, we consider solely
the effects from the scalar operator $O_\U$. 
The partonic cross section  can be derived as
\begin{equation} 
\frac{ d^2 \hat \sigma}{d \hat t d P_\U^2}
 = \frac{1}{16\pi \hat s^2} 
|\overline{\cal M}|^2 \,
 \frac{1}{2\pi} \, A_{d_{\cal U}} \, 
 \left( \frac{P_\U^2}{\Lambda_\U^2} \right)^{d_{\cal U}-2} \,
  \frac{1}{\Lambda_\U^2} 
\end{equation}
with the following matrix element squared for  subprocesses 
\begin{equation}
 |\overline {\cal M}(gg \to g{\cal U})|^2 =
\frac{ 1536\pi\alpha_s}{4\cdot 8\cdot 8} \,
\lambda_0^2 \frac{ (P_\U^2)^4 + \hat s^4+ \hat t^4+ \hat u^4}
{\hat s  \hat t\hat u \Lambda^2_{\U} } \; ,
\end{equation}
\begin{equation}
   |\overline {\cal M}(q\bar q \to g{\cal U}) |^2 
  = \ \frac{8}{9} g_s^2 \lambda_1^2 \,
\frac{ (\hat t-P_\U^2)^2 + (\hat u - P_\U^2)^2} {\hat t \hat u} \; ,
\end{equation}
\begin{equation}
  |\overline {\cal M}(qg \to q{\cal U})|^2 
 =  - \frac{1}{3} g_s^2 \lambda_1^2  \,
\frac{ (\hat t-P_\U^2 )^2 + (\hat s - P_\U^2)^2 }{\hat s \hat t} \; ,
\end{equation}
and a formula similar to the last one applies for 
$\bar qg \to \bar q{\cal U}$ as well.
Note that the gluon fusion process involving  $\lambda_0$ is further
suppressed by dimension counting.
%%%%
Although $P_\U^2$ is related to $\hat s$ by a kinematic relation
similar to
Eq.~(\ref{kinematic}), 
it is not uniquely determined at hadronic level
where $\hat s \sim x_1 x_2 s$ with $s$ the center-of-mass energy
squared of the colliding hadrons and $x_{1,2}$ are the parton momentum
fractions.  
We found that the peculiar feature of the phase space factor
$A_{d_\U}$ as a function of $d_\U$ at partonic level is more or less
washed out.  With only one jet in the final state, not many observables
can be constructed.  We show in Fig. \ref{monojet} the energy spectrum
of the monojet at the LHC.  Since the $\hat s$ of each collision is unknown
due to parton smearing, the $P^2_\U$ of each event cannot be reconstructed.
%The peculiar feature of unparticle can hardly be seen in the energy spectrum.
%%
Therefore, it would be difficult to detect the unparticle
at hadronic environment using the mono-jet signal, in contrast to
its original anticipation \cite{georgi1}.
One would anticipate that mono-photon or mono-$Z$ production 
plus an unparticle may be more promising at hadronic collisions, because of
better experimental resolution for photons and charged leptons.  However,
one still suffers from the unknown $\hat s$ in hadronic collisions.  The
unparticle information carried by the mono-photon or mono-$Z$ is likely
to be washed out by parton smearing as well. 
Even though we do not consider the case of spin-2 unparticle here, including 
them should not alter the conclusion.

\subsection{Present constraints on $\Lambda_\U$ from mono-photon
production at LEP2}

\begin{figure}
\centering
\includegraphics[width=4.5in]{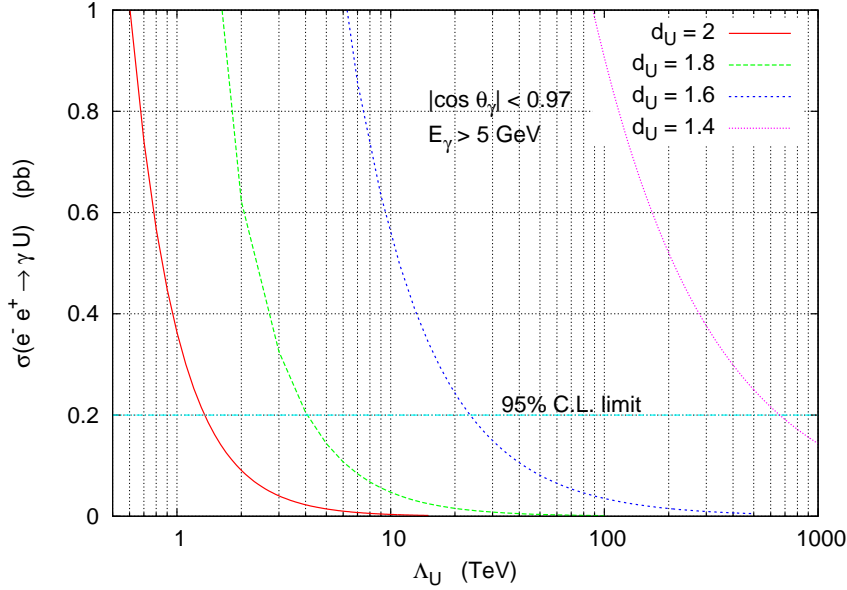}
\caption{\small \label{eegU-limit}
Cross sections for mono-photon plus unparticle production at 
the $e^- e^+$ collider with $\sqrt{s} = 207$ GeV for 
$d_\U =$ 1.4, 1,6, 1.8 and  2. 
The horizontal line of 0.2 pb is the 95\% C.L. upper limit.
}
\end{figure}

LEP collaborations \cite{lep-ph}
had measured mono-photon production in the 
context of extra dimensions, gauge-mediated SUSY breaking models, and other
models that can produce a single photon plus missing energy
in the final state.  Their limits on mono-photon production are similar.
We simply take the strongest bound 
among these LEP results: L3 obtained an 95\% C.L.
upper limit on $\sigma ( e^- e^+ \to \gamma + X) \simeq 0.2$ pb under
the cuts: $E_\gamma > 5$ GeV and $|\cos \theta_\gamma |< 0.97$ at
$\sqrt{s} = 207$ GeV.
We calculate mono-photon plus unparticle production with the same cuts
in $e^- e^+$ collisions with $\sqrt{s}=207$ GeV versus the unparticle
scale $\Lambda_\U$ (with a fixed $\lambda_1 = 1$) for 
$d_\U = $ 1.4, 1.6, 1.8 and 2 in 
Fig. \ref{eegU-limit}.  We have also drawn the horizontal line showing
the 95\% C.L. upper limit (0.2 pb).  The limits on $\Lambda_\U$ can be
read off where the horizontal line intercepts the curves.  We tabulate the 
limits in Table \ref{table1}. 
Since the production cross section scales as 
$\lambda_1^2/ \Lambda_\U^{2d_\U - 2}$, the limits increases very rapidly
when $d_U$ decreases from 2 to 1.4 with $\lambda_1$ fixed.  

\begin{table}[ht!]
\centering
\caption{\small \label{table1}
Limits on $\Lambda_\U$ from mono-photon
production data of 
$\sigma(e^- e^+ \to \gamma+X) \simeq 0.2$ pb at LEP2 (95\% C.L.)} 
\begin{tabular}{cc}
\hline
$d_\U$   &   $\;\; \Lambda_\U$ (TeV)  \\
\hline
$2.0$    &   1.35 \\
$1.8$    &   4   \\ 
$1.6$    &   23 \\
$1.4$    &   660 \\
\hline
\end{tabular}
\end{table}

\subsection{Other real emission processes}

The first operator in Eq.~(\ref{lambda2}) can involve the left-handed
lepton or quark doublet.  Therefore, it can give rise to $Z \to \nu
\bar \nu \U$ and charged-current process such as $W^- \to \ell^- \bar
\nu \U$ etc.  These decays will affect the invisible width of the $Z$
boson and the missing energy spectrum of the charged $W$ boson decay.
Analysis of LEP data for these decays could provide useful constraints on the 
scale of unparticle physics.

\section{Phenomenology: Virtual Exchanges at tree level}

\subsection{ Drell-Yan process}

Since the spin-0 operators often bring in a factor proportional to the 
external light
fermion mass in the amplitude,  their contributions are in general very small.
Here we only consider the contributions from spin-1 and spin-2 
unparticle exchange to the Drell-Yan process.

\subsubsection{Spin-1 unparticle}
\begin{figure}[t!]
\includegraphics[width=4.5in]{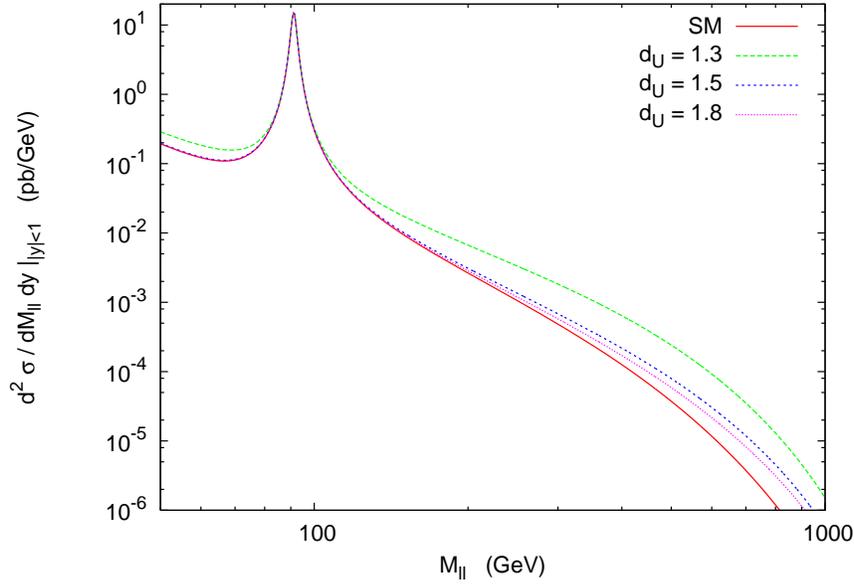}
\caption{
\label{drell-yan} \small
Drell-Yan invariant mass distribution for $d_{\cal U}=1.3,\,1.5$ and 
1.8 at the Tevatron with $\sqrt{s} = 1.96$ TeV.  We have chosen 
$\Lambda_\U = 1$ TeV and 
$\lambda_1 = 1$ for illustration.
}
\end{figure}

The effect of including the spin-1 unparticle virtual exchange in the 
Drell-Yan process 
has been studied in Ref. \cite{ours}.  We include here for completeness.
The differential cross section for the Drell-Yan process can be written as
\begin{equation}
\frac{d^2 \sigma}{ d M_{\ell\ell} \, dy}  =  K
  \frac{M^3_{\ell\ell}}{72\pi s}\,
 \sum_q \, f_q(x_1) f_{\bar q}(x_2) \;
 \times \left (
  | M_{LL} |^2 +   | M_{LR} |^2  +  | M_{RL} |^2  +  | M_{RR} |^2 
             \right ) \,,
\end{equation}
where $\hat s = M^2_{\ell\ell}$ and $\sqrt{s}$ is the center-of-mass
energy of the colliding hadrons.  $M_{\ell\ell}$ and $y$ are
the invariant mass and the rapidity of the lepton pair, respectively,
and $x_{1,2} = M_{\ell\ell}e^{\pm y}/\sqrt{s}$.
The $K$ factor equals $1 + \frac{\alpha_s}{2\pi} \frac{4}{3} \left(
  1+ \frac{4 \pi^2}{3} \right )$.
The reduced amplitude $M_{\alpha\beta} (\alpha,\beta = L,R)$ is given
by
\begin{equation}
 M_{\alpha \beta}  =  
  \     \lambda_1^2 Z_{d_\U}  \frac{1}{\Lambda_\U^2} 
 \left (- \frac{\hat s}{\Lambda_\U^2} \right)^{d_\U-2} 
+ \frac{ e^2 Q_l Q_q}{ \hat s}
  + \frac{e^2 g^l_\alpha g^q_\beta}{ \sin^2 \theta_{\rm w}
  \cos^2\theta_{\rm w}
 }\, \frac{1}{\hat s - M_Z^2+iM_Z\Gamma_Z} \; .
\label{reduced}
\end{equation}
%
%%%
Since $\hat s > 0$, the phase factor $\exp(-i \pi d_\U )$ in the unparticle 
4-fermion contact term will interfere with the photon and $Z$ boson propagator 
in a rather non-trivial way. This unparticle propagator phase can 
interfere with 
both the real photon propagator as well as
the real and imaginary parts of the unstable  $Z$ boson propagator.  
This gives rise to interesting interference patterns \cite{georgi2}. 
Despite having a complex phase in the unparticle propagator,
it has been demonstrated in \cite{Stephanov} using deconstruction that 
this doesn't lead to an unstable unparticle.
As mentioned earlier, we can allow different
couplings in different chirality combinations in the 4-fermion contact
interactions. In fact, we are able to
reproduce the effects in Ref. \cite{georgi2} using our 4-fermion amplitudes
with different chirality couplings.  However, it may be difficult to
disentangle the fractional differences from the SM prediction
in Drell-Yan production due to experimental uncertainties.  
It may be easier to test the angular distributions and 
interference patterns in $e^- e^+$ collisions.  We will show the results
in the next subsection.
For the moment we assume the same coupling in different chirality 
combinations so that the 4-fermion interactions are vector-like.
In Fig.~\ref{drell-yan}, we depict the Drell-Yan distribution as
a function of the invariant mass of the lepton pair for various 
$d_\U$ at the Tevatron. The peculiar effects from the phase space factor of
$A_{d_\U}$ for non-integral values of $d_\U$ are evident.

\subsubsection{Spin-2 unparticle}
We also study the effect of exchanging a spin-2 unparticle in Drell-Yan
process with the first operator in Eq.~(\ref{lambda2}).  
Similar pursuit has been performed in \cite{Mathews-Ravindran}.
The amplitude
for $q (p_1) \bar q (p_2) \to e^- (p_3) e^+ (p_4)$
due to unparticle exchange can be adapted from Eq.~(\ref{4-fermion-spin2})
with the substitutions $p_2 \to - p_2$ and $p_4 \to - p_4$:
\begin{eqnarray}
i {\cal M}_\U &=& -\frac{i}{8}  \lambda_2^2 \, Z_{d_\U} 
\, \frac{1}{\Lambda_\U^4} \, 
 \left( - \frac{ \hat s }{\Lambda_\U^2} \right)^{d_\U - 2}  \,
 \biggr[ (p_1 - p_2 ) \cdot (p_3 - p_4)\, 
\bar v(p_2) \gamma^\mu u(p_1) \, \bar u(p_3) \gamma_\mu v(p_4) \nonumber \\
&& + \; \bar v(p_2) \left( \not\!{p}_3 - \not\!{p}_4 \right ) u(p_1) \,
  \bar u(p_3) \left( \not\!{p}_1 - \not\!{p}_2 \right ) v(p_4) \,
   \biggr ] \;.
\end{eqnarray}
Let us write the constant pre-factor in $i {\cal M}_\U$ as
\[
  A =  - \frac{1}{8}\,\lambda_2^2 \, Z_{d_\U} 
\, \frac{1}{\Lambda_\U^4} \, 
 \left( - \frac{ \hat s }{\Lambda_\U^2} \right)^{d_\U - 2}
\]
which includes the unparticle phase $\exp (-i \pi d_\U )$ for $\hat s > 0$.
The complete amplitude squared without color- or spin-average is given by
\begin{eqnarray}
 \sum |{\cal M}|^2 &=& \Biggr \{
  4 \hat u^2 \left( | M^{\rm sm}_{LL} |^2 + |M^{\rm sm}_{RR}|^2 \right ) +
  4 \hat t^2 \left( | M^{\rm sm}_{LR} |^2 + |M^{\rm sm}_{RL}|^2 \right )
  \nonumber \\
&+& 8 |A|^2 \, \left( 
 \hat t^4 + \hat u^4 - 6 \hat t^3 \hat u - 6 \hat t \hat u^3
                 + 18 \hat t^2 \hat u^2 \right ) 
+ 16 \frac{e^2 Q_e Q_q}{\hat s} \, \Re \mathrm{e} (A) \, (\hat u - \hat t)^3
  \nonumber \\
&+& 16 \frac{e^2}{\sin^2\theta_{\rm w} \cos^2 \theta_{\rm w} }
 \Re \mathrm{e} \left( \frac{A^*}{ \hat s - M_Z^2 + i M_Z \Gamma_Z} \right )
\, \biggr[ g_a^e g_a^q 
 \left( \hat t^3 - 3 \hat t^2 \hat u - 3 \hat t \hat u^2 + \hat u^3 \right )
                   \nonumber \\
    && + \;    \Biggl. g_v^e g_v^q (\hat u- \hat t)^3 \biggr]  \Biggr\} \;,
\label{spin2}
\end{eqnarray}
where
\begin{eqnarray}
 M^{\rm sm}_{\alpha \beta}  &=&  
\frac{ e^2 Q_l Q_q}{ \hat s}
  + \frac{e^2 g^l_\alpha g^q_\beta}{ \sin^2 \theta_{\rm w}
  \cos^2\theta_{\rm w}
 }\, \frac{1}{\hat s - M_Z^2+iM_Z\Gamma_Z} \;, \;\;\;\;
   \alpha,\beta = L,R \nonumber \\
 g_v^f &=& \frac{g_L^f + g_R^f}{2} \;, \nonumber \\
 g_a^f &=& \frac{g_L^f - g_R^f}{2} \; .\nonumber 
\end{eqnarray}
The differential cross section for the subprocess is 
\begin{equation}
\frac{ d \hat \sigma}{d \cos \theta^*} (q\bar q \to e^- e^+) =
\frac{1}{32 \pi \hat s}\, \left( \frac{1}{3} \frac{1}{4} \sum |{\cal M}|^2
 \right ) \;,
\end{equation}
where $\theta^*$ is the scattering angle in the parton rest frame, and
$\hat t = - \frac{\hat s}{2} ( 1- \cos\theta^* )$,
$\hat u = - \frac{\hat s}{2} ( 1 + \cos\theta^* )$, and the factor 
$\frac{1}{3} \frac{1}{4}$ is for the color and spin average of the 
initial partons.
Integrating over $\cos\theta^*$ from $-1$ to $1$, the subprocess cross section
is 
\begin{equation}
\hat \sigma (q\bar q \to e^- e^+) = \frac{1}{144 \pi \hat s} \biggr [
  \hat s^2 \left( |M^{\rm sm}_{LL}|^2 + |M^{\rm sm}_{RR}|^2
                + |M^{\rm sm}_{LR}|^2 + |M^{\rm sm}_{RL}|^2 \right )
 + \frac{12}{5} |A|^2 \hat s^4 \biggr ] \;.
\end{equation}
It is noted that once when $\cos\theta^*$ is integrated,
the interference term goes to zero accidentally.  Therefore, it is hard
to discriminate the effect of spin-2 unparticle by the invariant mass spectrum
because of high suppression of powers of $\Lambda_{\U}$ in the
quantity $A$.  
Only the angular distribution can show a discernible effect, but the angular
distribution is somewhat smeared out in Drell-Yan production because 
the central scattering angle is boosted by the partons.

There is another contribution from the subprocess $gg \to \U^* \to e^- e^+$
via a tree-level exchange of a spin-2 unparticle.
Such a possibility arises from both operators in Eq.~(\ref{lambda2}) in which
we assume they have the same couplings.
The spin- and color-averaged amplitude squared for this process is given by
\begin{equation}
 |{\cal \overline M}|^2 ( gg \to e^- e^+) 
 = 4 |A|^2 \hat u \hat t ( \hat u^2 + \hat t^2 ) \,.
\end{equation}
The integrated subprocess cross section is 
\begin{equation}
\hat \sigma ( gg \to e^- e^+) 
 = \frac{1}{40\pi} |A|^2  \hat s^3 \;.
\end{equation}
Folded with parton distribution functions we obtain
\begin{eqnarray}
\frac{d^2 \sigma}{ d M_{\ell\ell} \, dy}  & = &  K
  \frac{1}{72\pi s}\, \Biggr \{
 \sum_q \, f_q(x_1) f_{\bar q}(x_2) \nonumber\\
& & \times \biggr [ M_{\ell\ell}^3 
\left ( 
  | M^{\rm sm}_{LL} |^2 +   | M^{\rm sm}_{LR} |^2  +  | M^{\rm sm}_{RL} |^2
  +  | M^{\rm sm}_{RR} |^2  \right ) 
+ \frac{12}{5} M_{\ell\ell}^7 \,| A|^2 \biggr ]  \nonumber \\
&+& 
  f_g(x_1) f_g(x_2)\, \frac{18}{5}\, M^7_{\ell\ell} \,| A|^2 \Biggr \} \,.
\end{eqnarray}
It is clear that the invariant mass distribution depends on $|A|^2$ rather 
than 
linear in $A$.  Therefore, it needs a rather large coupling for the unparticle 
operator in order to see the effect, given a large $\Lambda_{\U}$.  
We do not intend to show the invariant mass distribution here because it does
 not 
have any special feature.  One would rather attempt to look at the angular 
distribution, which has a linear dependence on $A$.  However, at hadronic 
machines one has to boost back to the rest frame of the lepton pair in order
to obtain the scattering angle.  Thus, experimental uncertainties are involved.
We would turn to the study of the angular distributions in fermion-pair
production at $e^- e^+$ colliders, which is more direct and 
the center-of-mass energy of the collision is uniquely specified.

\subsection{Fermion-pair production at $e^- e^+$ colliders}
The fermion pair production at $e^- e^+$ colliders can be studied using
the amplitude in Eq.~(\ref{reduced}) and the  amplitude squared 
in Eq.~(\ref{spin2}) with appropriate color-factor modifications
for spin-1 and spin-2 unparticle exchange, respectively. 

\subsubsection{Spin 1 unparticle}
The differential cross section including the spin-1 unparticle 
exchange is given by
\begin{equation}
\frac{d\sigma (e^- e^+ \to f\bar f)}{d \cos\theta} =
  \frac{N_c s}{128 \pi} \left[ ( 1+ \cos\theta)^2 ( |M_{LL}|^2 + |M_{RR}|^2 )
                             + ( 1- \cos\theta)^2 ( |M_{LR}|^2 + |M_{RL}|^2 ) 
                        \right ] \;,
\label{ffeespin1}
\end{equation}
where $M_{\alpha\beta}$'s are given by Eq.~(\ref{reduced}).

\begin{figure}[t!]
\centering
\includegraphics[width=3.2in]{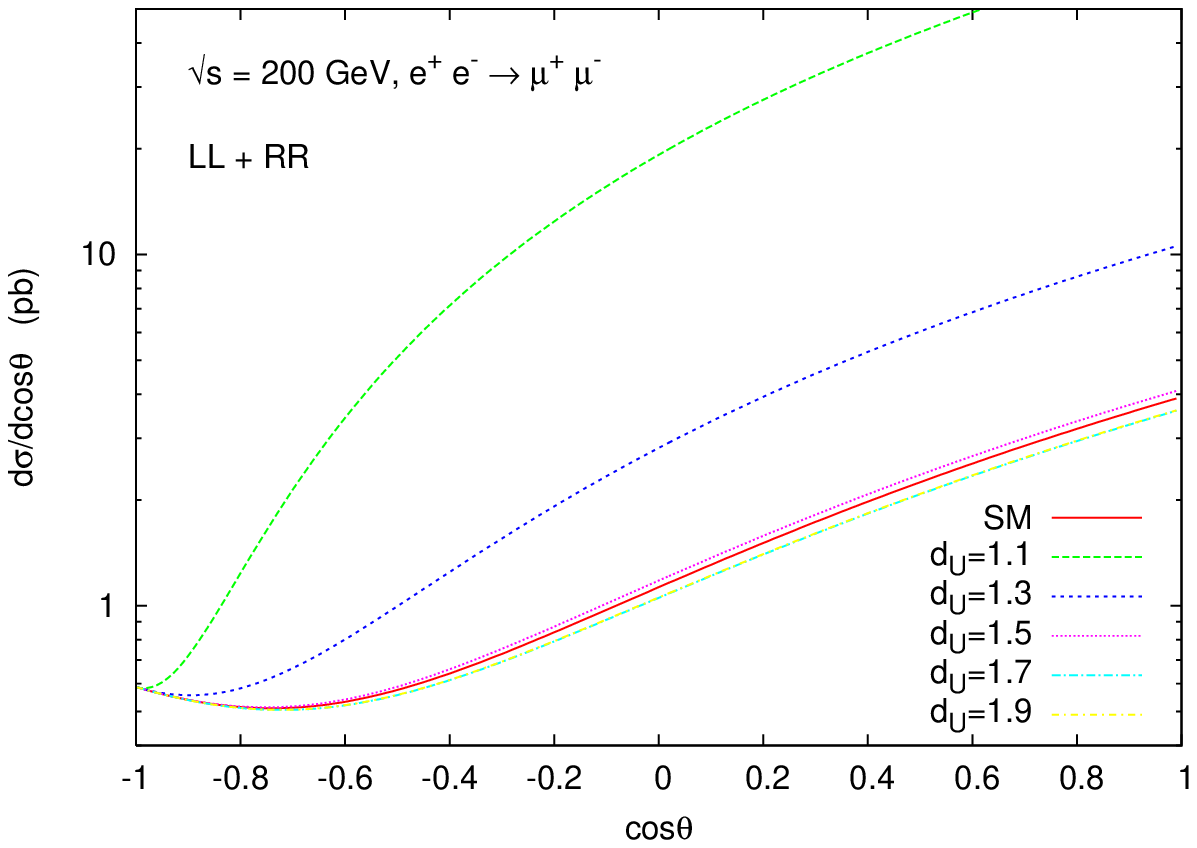}
\includegraphics[width=3.2in]{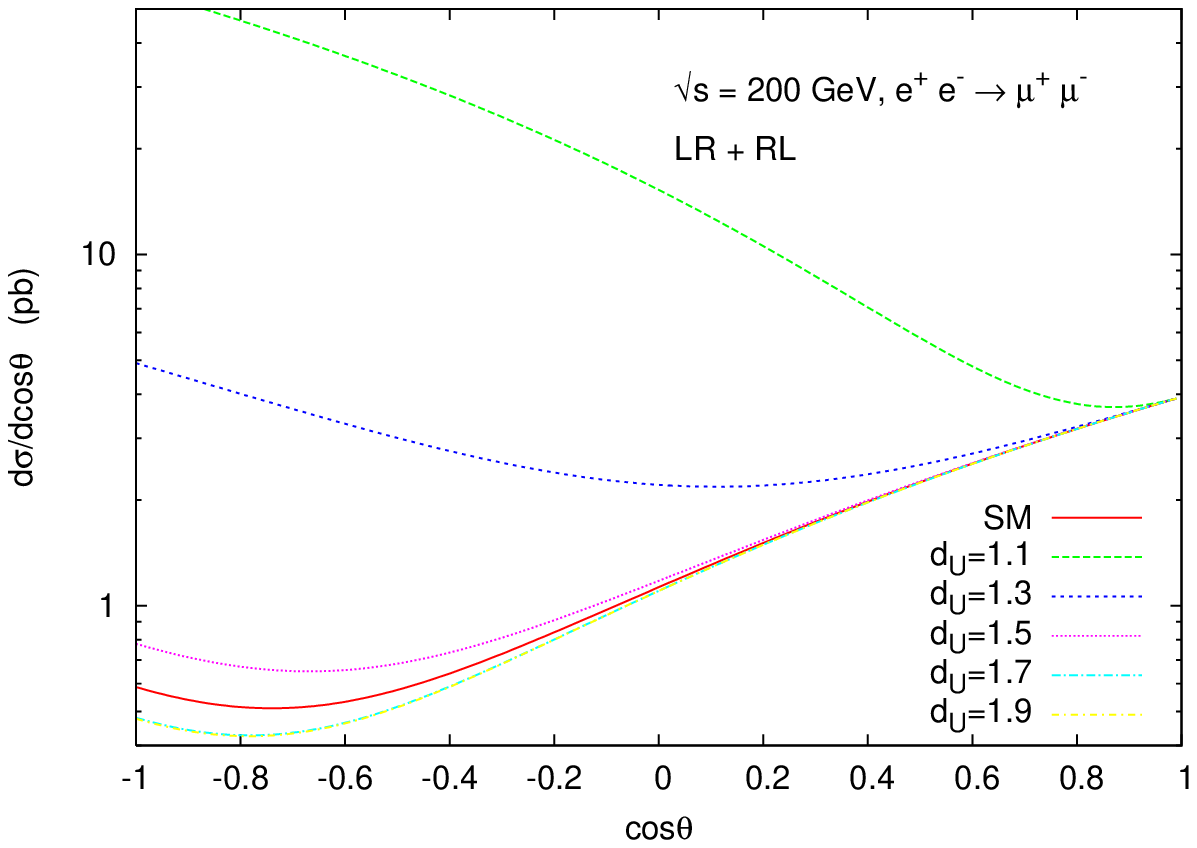}
\caption{\small \label{ee-ff-spin1-ang}
Angular distributions for $e^-e^+ \to \mu^- \mu^+$ with various $d_\U$
at $\sqrt{s} = 200$ GeV.  The left (right) panel is with $LL+RR$ ($LR+RL$) 
contact terms
plus the SM contributions. We have set $\Lambda_{\U} = 1$ TeV and 
$\lambda_1 = 1$. }
\end{figure}

To reiterate, the unparticle 4-fermion contact interactions in 
Eq.~(\ref{4fermionsop})
can be different for different chiralities of the fermions.  Let us write
the contact term between an electron and a fermion $f$ as
\begin{eqnarray}
{\cal M}_1^{ef} = \lambda_1^2 \, Z_{d_\U} \,\frac{1}{\Lambda_\U^2}
 \,  \left( - \frac{P_\U^2}{\Lambda_\U^2}  \right)^{d_\U - 2} 
\sum_{\alpha, \beta =L,R} \, \eta_{\alpha\beta}
( \bar e \gamma_\mu P_\alpha e)\,  ( \bar f \gamma^\mu P_\beta f) \;,
\label{4f}
\end{eqnarray}
where $P_{L,R} = (1 \mp \gamma_5)/2$ are the chirality projection operators,
and $\eta_{\alpha\beta}= \pm 1, 0$.
It is clear from Eq.~(\ref{ffeespin1}) that different modifications to 
$M_{\alpha\beta}$ can significantly change the angular distribution, because 
$M_{LL}$ and $M_{RR}$ are multiplied by $(1+\cos\theta)^2$ while 
$M_{LR}$ and $M_{RL}$ are multiplied by $(1-\cos\theta)^2$.
We show in Fig. \ref{ee-ff-spin1-ang} 
the angular distribution for $e^- e^+ \to \mu^- \mu^+$ at
$\sqrt{s}=200$ GeV, with (a) only $LL+RR$ and (b) only $LR+RL$ contact
interactions.  It is easy to understand why $LL+RR$ is increased in the positive
region of $\cos\theta$ while $LR+RL$ is enhanced in the negative
$\cos\theta$ region.  The forward-backward asymmetry can therefore discriminate
various chirality couplings.

\begin{figure}[t!]
\centering
\includegraphics[width=3.2in]{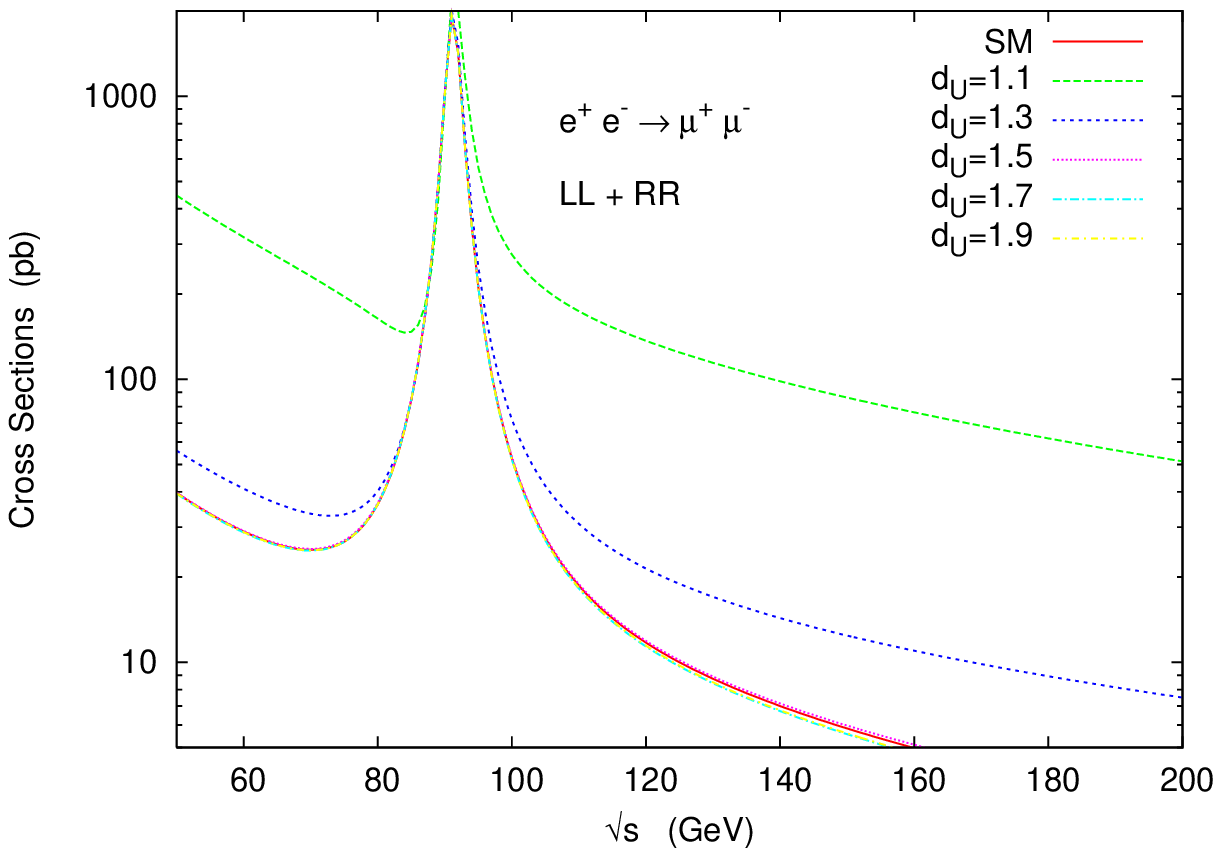}
\includegraphics[width=3.2in]{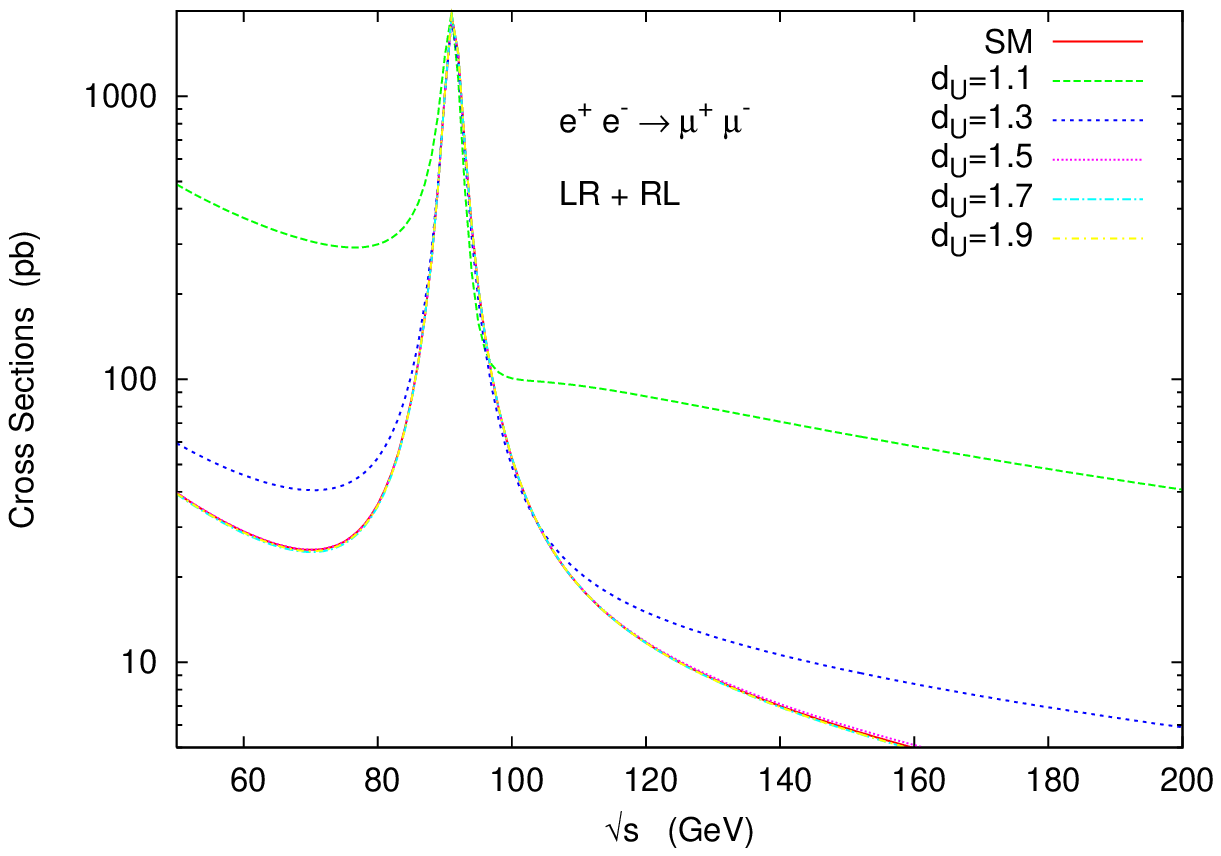}
\caption{\small \label{ee-ff-spin1}
Total cross sections for $e^- e^+ \to \mu^- \mu^+$ 
versus $\sqrt{s}$ with various $d_\U$.  
The left (right) panel is with $LL+RR$ ($LR+RL$) 
contact terms
plus the SM contributions.
We have set $\Lambda_{\U} = 1$ TeV and $\lambda_1 = 1$. 
}
\end{figure}

The integrated cross section for $e^- e^+ \to f \bar f$ can be obtained as
\begin{equation}
 \sigma (e^- e^+ \to f \bar f) =  \frac{N_c s}{48 \pi} \left (
  |M_{LL}|^2 + |M_{RR}|^2  +  |M_{LR}|^2 + |M_{RL}|^2 \right ) \,.
\end{equation}
As mentioned before when we calculated the 4-fermion contact interactions,
the unparticle propagator has a phase $\exp(-i \pi d_\U)$, which can 
interfere with the real and imaginary parts of the $Z$ boson propagator.
We show in Fig. \ref{ee-ff-spin1} the total cross sections for
$e^- e^+ \to \mu^- \mu^+$ versus $\sqrt{s}$ in the vicinity of the $Z$ pole,
with (a) $LL+RR$ contact terms and (b) $LR+RL$ contact terms.
Interesting interference patterns can be seen around the $Z$ pole.

\subsubsection{Spin 2 unparticle}
\begin{figure}[t!]
\includegraphics[width=4.5in]{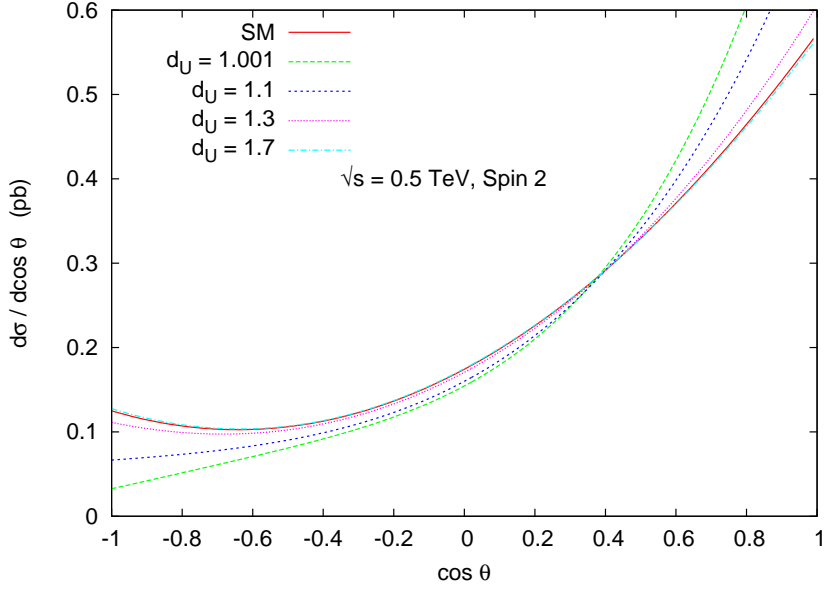}
\caption{
\label{ee-ff-spin2} \small
Angle distribution for $e^- e^+ \to \mu^- \mu^+$ with 
spin-2 unparticle exchange plus SM contributions at $\sqrt s$ = 0.5 TeV. 
We have set $\Lambda_{\U} = 1$ TeV and $\lambda_2 = 1$.
}
\end{figure}

The differential cross section including the spin-2 unparticle 
exchange can be obtained as
\begin{equation}
\frac{d\sigma (e^- e^+ \to f\bar f)}{d \cos\theta} = 
  \frac{1}{32 \pi s } \left( N_c \frac{1}{4} \sum |{\cal M}|^2 \right ) \;,
\end{equation} 
where $\sum |{\cal M}|^2$ is given in Eq.~(\ref{spin2}). 
We show in Fig. \ref{ee-ff-spin2} the angular distribution for 
$e^- e^+ \to \mu^- \mu^+$ at $\sqrt{s} = 0.5$ TeV with various $d_\U$.  For 
$d_\U < 1.3$, features of spin-2 unparticle exchange can be easily seen.

Integrating over $\cos\theta$ from $-1$ to $1$, we obtain the total cross
section:
\begin{equation}
\sigma (e^- e^+ \to f\bar f) = 
  \frac{N_c}{48 \pi s } \biggr[ s^2 \left(
  | M^{\rm sm}_{LL} |^2 +   | M^{\rm sm}_{LR} |^2  +  | M^{\rm sm}_{RL} |^2
  +  | M^{\rm sm}_{RR} |^2  \right ) 
+ \frac{12}{5} s^4\, |A|^2 \biggr ] \;.
\end{equation} 
Similar to Drell-Yan production the interference term linearly proportional to
$A$ goes to zero accidentally.  Therefore, the total cross section is 
not a sensitive probe for spin-2 unparticle exchange.

\subsection{Diphoton production}
\begin{figure}[t!]
\includegraphics[width=4.5in]{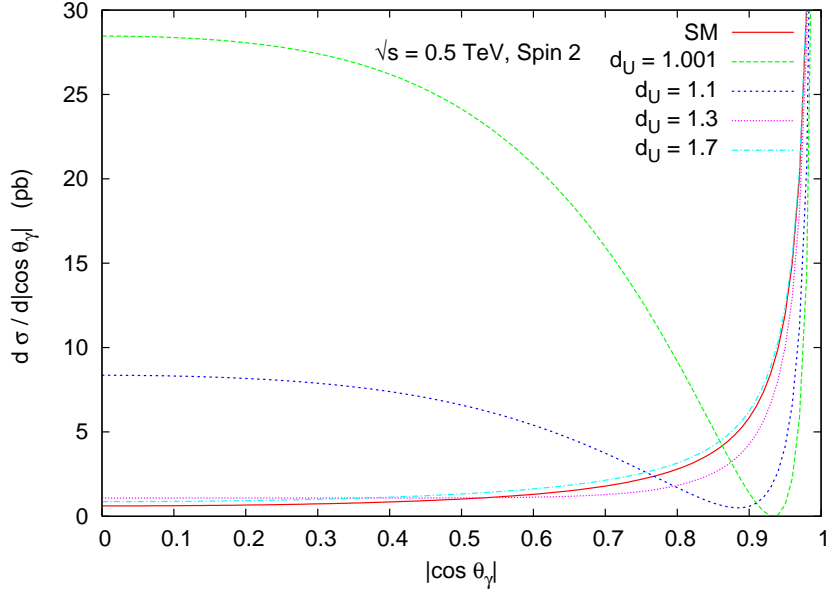}
\caption{
\label{eegg-spin2-lambda5} \small
The differential cross section 
$\frac{d \sigma}{d |\cos\theta_\gamma|} ( e^- e^+ \to \gamma\gamma)$
versus $|\cos\theta_\gamma|$ at $\sqrt{s} = 0.5$ TeV with a spin-2 unparticle
virtual exchange plus standard model contributions.  $\lambda_2$ is 
set at $5$ for visibility and $\Lambda_{\U} =1 $ TeV.}
\end{figure}

\begin{figure}[t!]
\includegraphics[width=4.5in]{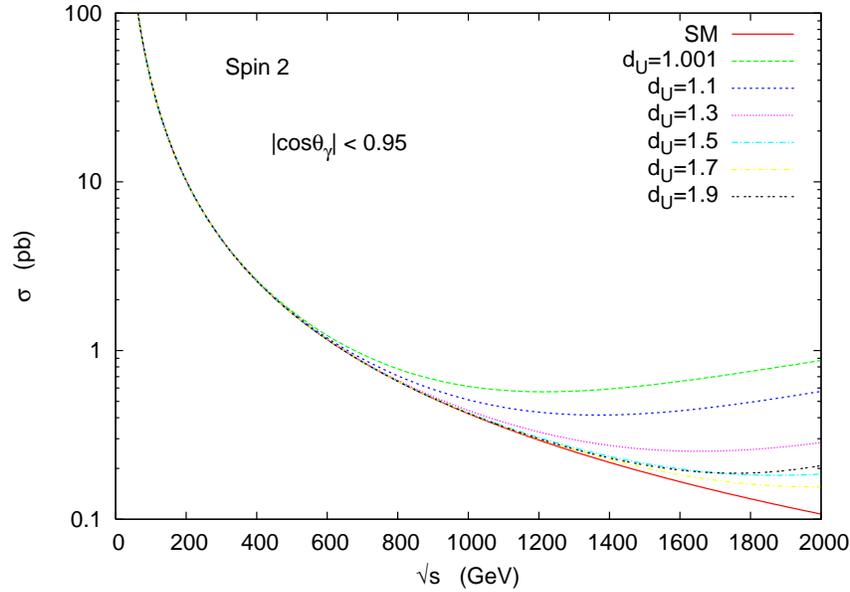}
\caption{
\label{eegg-total-lambda1} \small
Total cross section for $e^- e^+ \to \gamma \gamma$ with 
spin-2 unparticle exchange plus standard model contributions versus 
center-of-mass energy for different values of $d_\U$. 
We have set $\Lambda_{\U}=1$ TeV and 
$\lambda_2=1$.}
\end{figure}

Diphoton production at $e^- e^+$ and hadronic colliders have been proved 
very useful to detect unknown resonances that can decay into a pair of
photons and to search for anomalous diphoton couplings.  The spin-2 unparticle
can couple to a pair of fermions via the first operator of Eq.~(\ref{lambda2}) 
and to a pair of photons via the second operator in Eq.~(\ref{lambda2}).
There are three contributing Feynman diagrams: the $t$- and $u$-channel
standard model diagrams and the unparticle $s$-channel diagram.  The amplitude 
for $f(p_1) \,\bar f (p_2) \to \gamma (k_1) \, \gamma (k_2) $
due to the $s$-channel unparticle exchange is given by
\begin{eqnarray}
i {\cal M}_\U &=&  - \frac{i}{4}\, \lambda_2^2 \,Z_{d_\U}\, \left(
    \frac{ - s}{\Lambda_\U^2} \right )^{d_\U -2 }\, \frac{1}{\Lambda_\U^4}
\; \bar v(p_2) \left [ \gamma_\rho (p_1 - p_2)_\sigma + 
                       \gamma_\sigma (p_1 - p_2)_\rho \right ] u(p_1) 
\; \epsilon_\mu(k_1) \epsilon_\nu (k_2)
\nonumber \\
&\times & \biggr [ g^{\mu\nu} 
             \left( k_{1}^\rho k_2^\sigma +k_{2}^\rho k_1^\sigma \right )
                 + k_1 \cdot k_2 
       \left( g^{\rho\mu} g^{\sigma \nu}+g^{\sigma\mu}g^{\rho\nu}\right )
         \nonumber\\
& & \;\;\;\;\;\;\;\;\;\;\;\;\;\;\;\;\;\;\;\;\;\;\;\;\;\;\;\;\,
  - k_1^\nu \left(k_2^\rho g^{\sigma\mu} +k_2^\sigma g^{\rho\mu} \right )
 - k_2^\mu \left(k_1^\rho g^{\sigma\nu}+k_1^\sigma g^{\rho\nu} \right )
          \biggr ] \; .
\end{eqnarray}
Again, let us denote the constant pre-factor in $i{\cal M}_\U$ as
\begin{equation}
 A' = - \frac{1}{4}\, \lambda_2^2 \,Z_{d_\U}\, \left(
    \frac{ - s}{\Lambda_\U^2} \right )^{d_\U -2 }\, \frac{1}{\Lambda_\U^4}
\;.
\end{equation}
The spin- and 
color-averaged amplitude squared is given by
\begin{eqnarray}
 |{\cal \overline M}|^2 &=& 
 \frac{1}{4}\,\frac{1}{N_c}\, \biggr \{ 8 e^4 Q_f^4 
      \left ( \frac{u}{t}+\frac{t}{u} \right )  +  32 u t ( u^2 + t^2 ) | A'|^2
  + 32 e^2 Q_f^2 ( u^2 + t^2 ) \Re \mathrm{e} (A')  \biggr \} \,.
\end{eqnarray}
The differential cross section is given by
\begin{equation}
\frac{d \sigma}{d |\cos\theta_\gamma|} ( f \bar f \to \gamma\gamma)
= \frac{1}{32 \pi s}\,  |{\cal \overline M}|^2 \,, 
\end{equation}
where $ 0\le |\cos\theta_\gamma| \le 1$ because of identical photons in the
final state.
%%%
We show the angular distribution in Fig. \ref{eegg-spin2-lambda5}.  In the SM,
the angular distribution is very forward with majority of the cross section
at $|\cos\theta_\gamma|$ close to 1.  When $d_U$ is less than $1.2$ the 
majority comes  from the central region and a dip is formed around 
$|\cos\theta_\gamma| \approx 0.9$.  It is because of the spin-2 structure
of the operator.
The angular variable $|\cos\theta_\gamma|$ can be integrated from $0$ to
a cutoff $z$ because of the collinear divergence of the SM cross section
at $|\cos\theta_\gamma| = 1$.  We obtain the integrated cross section as
\begin{eqnarray}
\left. \sigma (f \bar f \to \gamma\gamma) \right |_{ 0 \le 
  |\cos\theta_\gamma| < z} &=& 
 \frac{1}{32\pi s} \, \frac{1}{4 N_c} \biggr \{
  8 e^4 Q_f^4 \left[ -2 z - 2 \log \left | \frac{1-z}{1+z} \right | \right ]
  \nonumber \\
&+& 32 s^4 \left( \frac{z}{8} - \frac{z^5}{40} \right ) |A'|^2
   +32 e^2 Q_f^2 s^2 \left( \frac{z}{2} + \frac{z^3}{6} \right ) \Re\mathrm{e}
(A') \biggr \} \; .
\end{eqnarray}
We show the total cross section of $e^- e^+ \to \gamma \gamma$ with a spin-2
unparticle exchange versus the center-of-mass energy in 
Fig. \ref{eegg-total-lambda1} with an angular cut of 
$|\cos\theta_\gamma| < 0.95$.  We have set $\Lambda_{\U}=1$ TeV and 
$\lambda_2=1$.  The cross section starts to show visible deviations
 when  $\sqrt{s}$ is around 0.5 TeV.

\subsection{Experimental constraints on unparticle scale $\Lambda_\U$}

Since the spin-1 unparticle exchanges will lead to 4-fermion contact
interactions, we can use the existing limits on 4-fermion contact
interactions \cite{contact-data},\cite{pdg} to constrain the unparticle scale
$\Lambda_\U$.  We can compare Eq.~(\ref{4f}) with the conventional 
4-fermion contact interactions
\begin{equation}
{\cal L}_{4f} = \frac{4 \pi}{\Lambda^2} 
  \sum_{\alpha, \beta =L,R} \, \eta_{\alpha\beta}
( \bar e \gamma_\mu P_\alpha e)\,  ( \bar f \gamma^\mu P_\beta f) \;,
\end{equation}
which results in the following equality:
\begin{equation}
\lambda_1^2 \, Z_{d_\U} \,\frac{1}{\Lambda_\U^2}
 \,  \left( - \frac{P_\U^2}{\Lambda_\U^2}  \right)^{d_\U - 2} 
  =  \frac{4 \pi}{\left( \Lambda^{95} \right)^2} \;,
\label{est}
\end{equation}
where $\Lambda^{95}$s are the 95\% C.L. limits on the $eeqq$ 
contact interaction scales obtained by combining global data on
fermion-pair production at LEP, Drell-Yan production at the Tevatron,
deep-inelastic scattering at HERA, and a number of low-energy 
parity-violating experiments \cite{contact-data}.

\begin{figure}[t!]
\centering
\includegraphics[width=4.5in]{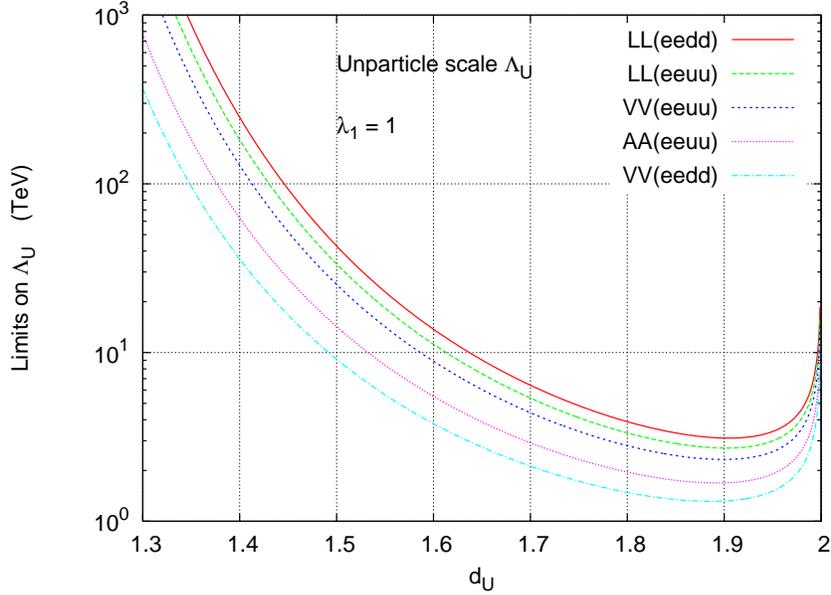}
\caption{ \small \label{limit-4f}
Rescaled limits from  existing 4-fermion contact interactions.  $LL$ means
only left-left chirality is considered while $VV$ means $LL+RR+LR+RL$
and $AA$ means $LL+RR-LR-RL$.  We have chosen $P_\U^2 \approx 
(0.2\; {\rm TeV} )^2$.
}
\end{figure}

Instead of performing a full analysis, we do a simple estimate here by
putting a fixed value for $P_\U^2$ into Eq.~(\ref{est}).  Since the 
limits are dominated by the LEP2 data \cite{contact-data} when
parity-conserving operators are considered, 
a fixed value of $P_\U^2 \approx (0.2 \;{\rm TeV})^2$ is chosen.  
Other choices are possible but will not affect our results significantly.
The best limit is on the $LL$ chirality because the parity-violating
experiments, especially the atomic-parity violation, are very stringent:
$\Lambda^{95}_{LL}(eeuu) \simeq  23 $ TeV while 
$\Lambda^{95}_{LL}(eedd) \simeq  26 $ TeV.  When parity-conserving
combinations are considered, the limits are lowered:
$\Lambda_{VV}^{95} (eeuu) \simeq 20$ TeV,
$\Lambda_{VV}^{95} (eedd) \simeq 12$ TeV, and
$\Lambda_{AA}^{95} (eedd) \simeq \Lambda_{VV}^{95} (eeuu) = 15$ TeV.
We rescale these 4-fermion contact interaction limits to the 
limits on the unparticle scale $\Lambda_{\U}$ using Eq.~(\ref{est}),
with $\lambda_1 =1$ and $P^2_{\U} = (0.2\;{\rm TeV})^2$.  The results
are shown in Fig. \ref{limit-4f}.  
Note that we have ignored the phase in the unparticle
propagator in the analysis.
The limits obtained are similar to those obtained from the single-photon
production at LEP2.  

The estimates here are rather crude, because we have substituted the
factor $P_\U^2$ by a constant $(0.2\;{\rm TeV})^2$, which should be good for 
a crude estimate.   In principle,
a different $P_\U^2$ is needed for analysis of each high energy process. 
An updated global analysis using $P_\U^2$ dependent amplitudes is necessary
for more accurate limits.  Similarly, another global analysis is needed 
for constraining the spin-2 unparticle exchange. 
We note that a recent paper 
\cite{Bander-Feng-Rajaraman-Shirman} has also derived some 
limits of the unparticle scale.

\section{Conclusions}

Scale invariance or the enlarged conformal invariance is an attractive
symmetry, but is not realized in the low energy visible world.  Perhaps, below a 
sufficient high energy scale an exact scale invariant hidden sector may exist.
Such a strictly scale invariant sector may couple weakly to the SM particles
such that we may be able to probe it via high energy processes at the
LHC and ILC.
Operators $\cal O_\U$ of a scale invariant sector with a general 
non-integral scale dimension $d_\U$ has a
phase space looked like a $d_{\cal U}$ number of invisible massless
particles.  Therefore, a typical reaction that involves 
emission of the unparticle in the final state gives rise to missing energy
signals in the detectors.
We have studied a number of processes that involve emission of the unparticle
in the final state, including $e^- e^+ \to \gamma \U,\; Z \U$ at the ILC
and $Z \to f \bar f \U$ at the $Z$-pole, as well as the monojet production
at the LHC.   We found that the energy distribution of the single photon
or the single $Z$ at ILC and the missing energy distribution in 
$Z\to f \bar f\U $ can discriminate the scale dimension $d_\U$.  However,
the monojet energy spectrum is not so sensitive to  $d_\U$
because of the wash-out by parton smearing.

We also formulate the virtual exchange of unparticles
between SM particles.  We have shown that spin-1 
unparticle exchange between two 
fermions gives rise to contact 4-fermion interactions, which 
scale as $(\hat s /\Lambda^2_{\U} )^{d_\U -1}$ and thus differ from
the conventional one because of the peculiar scale dimension $d_\U$.  
Spin-2 unparticle exchange gives rise to another form of 4-fermion
interactions.
We have used Drell-Yan production at hadronic colliders and fermion-pair
production at $e^- e^+$ colliders to study the interference of the 
unparticle-exchange amplitudes with the SM amplitudes.  One peculiar 
feature of unparticle propagator is the phase factor $\exp(-i\pi d_\U)$ 
which may interfere nontrivially with the $Z$ boson propagator.  We have
demonstrated the intriguing interference effects in great details 
in fermion-pair production 
in $e^- e^+$ collisions.
Finally, we have also studied diphoton production, which also shows 
the peculiar feature of the phase of unparticle propagator.  

Unparticles can be conjectured as a generalization of extra dimensions.
The number of extra dimensions only take on integral values while the scale dimension 
of unparticle can take on any, even non-integral values.  
We speculate on a relation $d_\U = n/2 + 1$ that relates the scale dimension 
to the number of large extra dimension.
Therefore, unparticle
physics is another program just as important as extra dimensions 
in the goals of the LHC.

Before we end, we offer a number of comments as follows.
\begin{enumerate}

\item The calculation of diphoton production can be easily extended to
other diboson production, such as $ZZ$ and $W^+ W^-$, at $e^- e^+$ 
and hadronic machines.  Likewise, one can study the unparticle effect
in the gauge boson scattering \cite{Greiner}.

\item The peculiar phase factor in the unparticle propagator can be used
as a strong phase that is required in the CP violation studies 
\cite{Chen-Geng} of the $B$-meson system.

\item It is more natural to assume that the unparticle sector is 
flavor blind.  
Flavor changing coupling of the SM particles with the unparticle 
can then be induced at 1-loop via $W$-boson exchange as was 
done in the second paper in \cite{Chen-Geng}.
Direct flavor changing couplings of the SM fermions  with unparticle
will suffer strong constraints from low-energy
flavor changing processes 
\cite{Luo-Zhu},
\cite{Chen-Geng},
\cite{Aliev-Cornell-Gaur},
\cite{Li-Wei},
\cite{Lu-Wang-Wang},
\cite{Choudhury-Ghosh-Mamta}. 
These constraints would push 
unparticle physics out of reach at the LHC.

\item Dijet production at hadronic colliders is also sensitive to unparticle
exchange.  It would be similar to diphoton production.  One would expect
enhancement of cross section at high invariant mass of the dijet.

\item Our formulas for 4-fermion contact interactions can be applied to
other areas, e.g., the $ep$ deep inelastic scattering \cite{Ding-Yan},
low-energy parity violating experiments,  $D - \bar D$ or $B -\bar B$ mixings
 \cite{Luo-Zhu,Chen-Geng,Li-Wei}, 
and atomic parity violation experiments \cite{contact}.

\item Quarkonium decays can also constrain the unparticle by their invisible 
widths and by the decay mode of $\gamma \; +$ nothing. 

\item Astrophysics places constraints on real emission of unparticles.  In 
principle, emission of unparticles in supernova, neutron stars, or some
other astrophysical systems can lead to substantial cooling other than
that by neutrinos.  Therefore, using the experimentally measured cooling
rates  one can constrain the unparticle scale. Various limits of the unparticle scale
have been estimated in \cite{Davoudiasl} from the supernova SN 1987A data as 
well as from other cosmological considerations.

\item The spin-1 unparticle contribution to the lepton anomalous
magnetic moment at 1-loop has been calculated \cite{ours}. It should
be possible to extend the calculation to the spin-2 case as well.  The
effect is expected to be minuscule, however.

\item Besides the 2-point function, the momentum part of a 3-point or
in general $n$-point function is known for a conformal field theory in
4 dimension up to an overall constant. Can one determine the overall
constant for the 3-point or in general $n$-point function for the
unparticle operators?  We will leave this to those with more ambitious
minds.

\end{enumerate}

Phenomenology of unparticle physics is quite rich. While the
underlying theory of unparticle is still needed to be unraveled by
theorists, experimentalists could detect such a hidden scale invariant
sector when the behemoth LHC machine becomes online in the year 2008!

\section*{Acknowledgments}
This research was supported in parts by the NSC
under Grant No.\ NSC 95-2112-M-007-001, NCTS
and U.S. DOE under Grant No. DE-FG02-84ER40173.

\appendix
\section{Functions $F$, $G$ and $H$}
These functions appeared in $f\bar f \to Z \U$ for spin-2 unparticle $\U$.
\begin{eqnarray}
(F,G,H) =(F_0,G_0,H_0)+\frac{1}{P^2_\U}(F_2,G_2,H_2) +\frac{1}{P^4_\U}(F_4,G_4,H_4) \nonumber
\end{eqnarray}
with
\begin{eqnarray}
F_0(t,u) & = & 2 t^2 u^2 
\left[ 
16 M_Z^6 + P^2_\U \left( 7 t^2 + 12 t u + 7 u^2 \right) - 3 \left( 3 t^3 + 11 t^2 u + 11 t u^2 + 3 u^3 \right) \right. \nonumber \\
& + & 6 M_Z^4 \left. \left( 7 P^2_\U - 2 \left( t + u \right) \right) + M_Z^2 \left( 14 P^4_\U 
-15 t^2 - 44 t u - 15 u^2 + 2 P^2_\U \left( t + u \right) \right) 
\right] \nonumber \\
G_0(t,u) & = & 
4 t u \left\{ 6 M_Z^6 \left( P_\U^2 - t - u \right) \left( t + u \right) \right.\nonumber\\
 &+& M_Z^4 \left[ 9 t^3 + 7 t^2 u + 7 
            t u^2 + 9 u^3 + 15 P_\U^4 \left( t + u \right) 
             -  2 P_\U^2 \left(12  t^2 + 19 t u + 12 u^2 \right) \right] \nonumber \\
             &+& t u  \left[ 6 P_\U^6 - 
              9 P_\U^4 \left( t +  u \right) - P_\U^2 \left( t^2 + 12 t u + u^2 \right) + 
               6 \left( t^3 + 6 t^2 u + 6 t  u^2 + u^3 \right) \right] \nonumber\\
      &+& M_Z^2 \left[ -3 t^4 + 
                    25 t^3 u + 58 t^2 u^2 + 25 
        t u^3 - 3 u^4 + 6 P_\U^6 \left( t +u \right) \right. \nonumber\\
               &-& P_\U^4 \left( 15 t^2 + 2 t u + 15 u^2 \right) + 2 P_\U^2 
            \left.   \left. \left( 6  t^3 - 11 t^2 u - 11 t u^2 + 6 u^3 \right) \right] \right\}
\nonumber\\
H_0(t,u) & = & 
24 M_Z^6 t u \left(-P_\U^2 + t + u \right)^2 \nonumber\\
&-& 6 M_Z^4 t u \left[ -9 P_\U^6 + 24 P_\U^4 \left(t + u\right) - P_\U^2 \left(21 t^2 + 38 t u + 21 u^2\right)  
 \right. \nonumber\\
&+& \left.  2 \left(3 t^3 + 5 t^2 u + 5 t u^2 + 3 u^3\right) \right] \nonumber\\
&-& M_Z^2 \left[  3 P_\U^8 
\left(  t^2 - 12 t u + u^2 \right) - 2 t u \left( t + u \right)^2 \left( 6 t^2 - 29 t  u + 6 u^2 \right) \right. \nonumber \\
    &-& 6 P_\U^6 \left( t^3 - 16 t^2 u - 16 t u^2 + u^3 \right) 
    + 54 P_\U^2 t u \left( t^3 + t^2 u + t u^2 + u^3 \right) 
    \nonumber\\
    &+&  \left. P_\U^4 \left( 3 t^4 - 102 t^3 u - 166 t^2 u^2 - 102 t u^3 + 3 u^4 \right) \right]
    \nonumber\\
   & + &
          t u \left[ 6 P_\U^{10} - 18 P_\U^8 \left(t + u\right) - 12 P_\U^4 \left(t + u\right)^3 
          + 3 P_\U^6 \left( 7 t^2 + 12 t u + 7 u^2 \right) 
          \right.
          \nonumber\\
      &-& \left. 18 
          t u \left( t^3 + 5 t^2 u + 5 t u^2 + u^3 \right) + P_\U^2 
          \left( 3 t^4 + 32 t^3 u + 78 t^2 u^2 + 32 t u^3 + 3 u^4 \right)
          \right]
\nonumber\\
F_2(t,u) & = &
2 t^2 u^2 \left( t + u \right) \left[ - 8 M_Z^4 \left( t + u \right) + 4 M_Z^2 \left( t^2 + 3 t u +  u ^2 \right)
+ 3 \left( t^3 + 5 t^2 u + 5 t u^2 + u^3 \right) \right] \nonumber \\
G_2(t,u) & = &
-4 t^2 u^2 \left( t + u \right) \left[-10  M_Z^4 \left(t + u\right) + 
      2 M_Z^2 \left( 3 t^2 + 7 t u + 3 u^2 \right) \right.
\nonumber \\
      && \;\;\;\;\;\;\;\; \;\;\;\;\;\;\;\; \;\;\;\;\;\;\;\; +3 \left. \left( t^3 + 5 t^2 u + 5 t u^2 + u^3 \right) \right]
\nonumber\\
H_2(t,u) & = &
2 t^2 u^2 \left( t + u \right)^2 \left[ -12 M_Z^4 + 8 M_Z^2 \left(t +  u\right) + 3 \left(t^2 + 4 t u + u^2\right) \right]
\nonumber\\
F_4(t,u) & = & H_4(t,u) = - \frac{1}{2} G_4(t,u)  \nonumber\\
&=&-2 t^2 u^2 \left( t + u \right)^3 \left( t^2 + u^2 - M_Z^2 \left( t + u \right) \right)
\nonumber
\end{eqnarray}
The following relations are found to be satisfied by these functions
\begin{eqnarray}
F_2 + G_2 + H_2 & = & 0 \; , \nonumber \\
F_4 + G_4 + H_4 & = & 0  \; . \nonumber 
\end{eqnarray}

%%%%%%%%%%%%%%%%%%%%%%%%%%%%%%%%%%%%%%%%

\end{document}